\documentclass[submission, Phys]{SciPost}

\hypersetup{
    colorlinks,
    linkcolor={red!50!black},
    citecolor={blue!50!black},
    urlcolor={blue!80!black}
}

\usepackage[bitstream-charter]{mathdesign}
\DeclareSymbolFont{usualmathcal}{OMS}{cmsy}{m}{n}
\DeclareSymbolFontAlphabet{\mathcal}{usualmathcal}

\newcommand\ket[1]{|#1\rangle}
\newcommand\bra[1]{\langle#1|}
\newcommand\tr{\operatorname{tr}}
\newcommand\imax{i_\text{max}}

\begin{document}

\begin{center}{\Large \textbf{Hilbert space fragmentation and slow dynamics in particle-conserving quantum East models\
}}\end{center}

\begin{center}
Pietro Brighi\textsuperscript{1$\star$},
Marko Ljubotina\textsuperscript{1} and
Maksym Serbyn\textsuperscript{1}
\end{center}

\begin{center}
{\bf 1} IST Austria, Am Campus 1, 3400 Klosterneuburg, Austria
\\
${}^\star$ {\small \sf pbrighi@ist.ac.at}
\end{center}

\begin{center}
\today
\end{center}


\section*{Abstract}
{\bf
Quantum kinetically constrained models have recently attracted significant attention due to their anomalous dynamics and  thermalization. In this work, we introduce a hitherto unexplored family of kinetically constrained models featuring conserved particle number and strong inversion-symmetry breaking due to facilitated hopping. We demonstrate that these models provide a generic example of so-called quantum Hilbert space fragmentation, that is manifested in disconnected sectors in the Hilbert space that are not apparent in the computational basis. Quantum Hilbert space fragmentation leads to an exponential in system size number of eigenstates with exactly zero entanglement entropy across several bipartite cuts. These eigenstates can be probed dynamically using quenches from simple initial product states. In addition, we study the particle spreading under unitary dynamics launched from the domain wall state, and find faster than diffusive dynamics at high particle densities, that crosses over into logarithmically slow relaxation at smaller densities. Using a classically simulable cellular automaton, we reproduce the logarithmic dynamics observed in the quantum case.  Our work suggests that particle conserving constrained models with inversion symmetry breaking realize so far unexplored universality classes of dynamics and invite their further theoretical and experimental studies.
}

\vspace{10pt}
\noindent\rule{\textwidth}{1pt}
\tableofcontents\thispagestyle{fancy}
\noindent\rule{\textwidth}{1pt}
\vspace{10pt}

\section{Introduction}
\label{sec:Intro}

In recent years, kinetically constrained models, originally introduced to describe classical glasses~\cite{Ritort2003,Garrahan2007,Fisher2004}, have received considerable attention in the context of non-equilibrium quantum dynamics~\cite{Garrahan2014,Garrahan2018a,Aidelsburger2021,Vasseur2021,Marino2022}. In analogy with their classical counterparts, they are characterized by unusual dynamical properties, including slow transport~\cite{Garrahan2018,Knap2020,Morningstar2020,Vasseur2021,Yang2022,Knap2022}, localization~\cite{Pancotti2020,Nandkishore2020,Pollmann2019} and fractonic excitations~\cite{Nandkishore2019a,Nandkishore2019c}. Additionally, in the quantum realm, other interesting phenomena have been observed, such as Hilbert space fragmentation~\cite{Nandkishore2020,Pollmann2020a,Pollmann2020b,Iadecola2020,Zadnik2021a,Zadnik2021b,Sen2021,Pozsgay2021} and quantum many-body scars~\cite{Michailidis2018,Choi2019,Papic2020,Iadecola2020a,Tamura2022}.

Among the many possible types of constraints, one can distinguish models that are inversion symmetric from those that break inversion symmetry.  Among the latter models, the so-called quantum East model~\cite{Garrahan2015,Pancotti2020,Lazarides2020,Marino2022,Garrahan2022,Eisinger1991} where spin dynamics of a given site is facilitated by the presence of a particular spin configuration \emph{on the left} represents one of the most studied examples. The quantum East model has been shown to host a localization-delocalization transition in its ground state~\cite{Pancotti2020}, which allows the approximate construction of excited eigenstates in matrix product state form. Transport in particle-conserving analogues of the East model was recently investigated through the analysis of the dynamics of infinite-temperature correlations, revealing subdiffusive behavior. A similar result has also been observed in spin-$1$ projector Hamiltonians~\cite{Pal2022}.

The interplay of particle conservation and kinetic constraints that break inversion symmetry opens several interesting avenues for further research. First, the phenomenon of so-called Hilbert space fragmentation that is known to occur in constrained models and is characterized by the emergence of exponentially many disconnected subsectors of the Hilbert space is expected to be modified. The additional $U(1)$ symmetry is expected to influence Hilbert space fragmentation beyond the picture presented in previously studied models~\cite{Garrahan2015,Pancotti2020,Marino2022}. Second, the presence of a conserved charge allows the study of transport~\cite{Vasseur2021,Yang2022,Knap2022}. While transport without restriction to a particular sector of fragmented Hilbert space results in slow subdiffusive  dynamics~\cite{Vasseur2021,Yang2022}, a recent work~\cite{Ljubotina2022} demonstrated that a restriction to a particular sector of fragmented Hilbert space can give rise to superdiffusion. This motivates the study of transport in the particle conserving East model restricted to a particular sector of the Hilbert space.

In this work, we investigate a generalized East model, consisting of hard-core bosons with constrained hopping. The constraint prevent hopping in the absence of bosons on a few preceding sites \emph{to the left}. The chiral nature of such facilitated hopping strongly breaks inversion symmetry, akin to the conventional East model, additionally featuring the conservation of the total number of bosons. Our results show that combining charge conservation and the breaking of inversion symmetry yield new interesting transport phenomena.
Specifically, we characterize the proposed generalized East model using its eigenstate properties and dynamics. 
The detailed study of the eigenstates reveals so-called quantum Hilbert space fragmentation, so far reported only in a few other models~\cite{Sen2021,Motrunic2022}. The quantum fragmentation we observe in our model leads to the existence of eigenstates that have zero entanglement along one or several bipartite cuts. The number of these low entanglement eigenstates increases exponentially with system size. We find that these unusual eigenstates can be constructed recursively, relying on special eigenstates existing in small chains that are determined analytically. Thus the particle-conserving East model provides an example of \emph{recursive quantum} Hibert space fragmentation.

The study of dynamics of the particle-conserving East model reveals that weakly entangled eigenstates existing in the spectrum can be probed by quenches from simple product states. In addition, the dynamics from a generic domain wall initial state reveals two distinct transport regimes. At short times dynamics is superdiffusive, whereas at longer times the constraint leads to a logarithmically slow spreading. We recover the logarithmically slow dynamics within a classically simulable cellular automaton that has the same features as the Hamiltonian model.  In contrast, the early time dynamical exponent differs between the quantum Hamiltonian dynamics and cellular automaton version. These results suggest that constrained models with inversion symmetry breaking and particle conservation may realize a new universality class of dynamics. This invites the systematic study of such models using large scale numerical methods and development of a hydrodynamic description of transport in such systems. 

The remainder of the paper is organized as follows. In Section~\ref{Sec:Model} we introduce the Hamiltonian of the particle-conserving East model and explain the effect of the constraint. We then investigate the nature of the Hilbert space fragmentation and of the eigenstates in Section~\ref{Sec:Eigenstates}. In Section~\ref{Sec:Dyn} we investigate the dynamical properties of the system, showing similarities in the long-time behavior among the quantum dynamics and the classical cellular automaton. Finally,  in Section~\ref{Sec:Disc}, we conclude by presenting a summary of our work and proposing possible future directions.

\section{Family of particle-conserving East models}\label{Sec:Model}

We introduce a family of particle conserving Hamiltonians inspired by the kinetically constrained East model in one dimension. The East model, studied both in the classical~\cite{Eisinger1991,Ritort2003} and quantum~\cite{Garrahan2015,Pancotti2020,Garrahan2022} cases, features a constraint that strongly violates inversion symmetry: a given spin is able to flip only if its \emph{left} neighbor is in the up ($\uparrow$) state. A natural implementation of such a constrained kinematic term in the particle-conserving case is a hopping process \emph{facilitated} by the presence of other particles on the left. The simplest example of such a model is provided by the following Hamiltonian operating on a chain of hard-core bosons, 
\begin{equation}
\label{Eq:Hr1}
\hat{H}_{r=1} = \sum_{i=2}^{L-1} \hat n_{i-1}\bigr(\hat{c}^\dagger_{i}\hat{c}_{i+1}+\hat{c}^\dagger_{i+1}\hat{c}_{i}\bigr),
\end{equation}
where the operator $\hat n_i =\hat{c}^\dagger_{i}\hat{c}_{i}$ is a projector onto the occupied state of site $i$. We assume open boundary conditions here and throughout this work, and typically initialize, without loss of generality, the first site as being occupied by a frozen particle. All sites to the left of the leftmost particle, in fact, cannot be occupied, hence they are not relevant to the behavior of the system. 

The Hamiltonian~(\ref{Eq:Hr1}) implements hopping facilitated by the \emph{nearest neighbor}  particle on the left, hence we refer to it as the range-1, $r=1$, particle conserving East model.  A natural extension of this model would be hopping facilitated by the nearest \emph{or} next nearest neighbor, which reads: 
\begin{equation}
\label{Eq:Hr=2}
\hat{H}_{2} = \sum_{i=2}^{L-1}(\hat{n}_{i-2}+\hat{n}_{i-1}-\hat{n}_{i-2}\hat{n}_{i-1})(\hat{c}^\dagger_{i}\hat{c}_{i+1} + \text{H.c.}), 
\end{equation}
where we treat the operator $\hat{n}_{i=0}=0$ as being identically zero. Note, that in this Hamiltonian we use the same hopping strength irrespective if the facilitating particle is located on the nearest neighbor or next nearest neighbor site, however this condition may be relaxed. Examples of range-1, $\hat H_1$,  and range-2, $\hat H_2$,  particle conserving East models can be further generalized to arbitrary range $r$ as 
\begin{eqnarray}
\label{Eq:Hr}
\hat{H}_r &=& \sum_{i=r+1}^{L-1} \hat{\mathcal{K}}_{i,r}\bigr(\hat{c}^\dagger_{i+1}\hat{c}_i+\text{H.c.}\bigr),
\\
\label{Eq:Kr}
\hat{\mathcal{K}}_{i,r} &=& \sum_{\ell=1}^{r}t_\ell\hat{\mathcal{P}}_{i,\ell}
\end{eqnarray}
where the operator $\hat{\mathcal{K}}_{i,r}$ implements a range-$r$ constraint using projectors on the configurations with $\hat{n}_{i-\ell}=1$ and the region $[i-\ell+1,i-1]$ empty, $\hat{\mathcal{P}}_{i,\ell} = \hat{n}_{i-\ell} \prod_{j=i-\ell+1}^{i-1} (1-\hat{n}_j)$. The coefficients $t_\ell$ correspond to the amplitude of the hopping facilitated by the particle located $\ell$-sites on the left. The Hamiltonian~$\hat{H}_2$ in Eq.~(\ref{Eq:Hr=2}) corresponds to the particular case when all $t_\ell=1$. 

Models with similar facilitated hopping terms were considered in the literature earlier.
In particular a pair hopping $\bullet\bullet\circ\leftrightarrow\circ\bullet\bullet$ was introduced in~\cite{DeRoeck2016} and later used in~\cite{Brighi2020} to probe many-body mobility edges, and shown to be integrable in Ref.~\cite{Pozsgay2021}.
In~\cite{Bahovadinov2022} a similar constrained hopping term was shown to arise from the Jordan-Wigner transformation of a next nearest neighbor XY spin chain. 
Another constrained model recently studied is the so-called \textit{folded} XXZ~\cite{Pollmann2019,Iadecola2020}, where the $\Delta\to \infty$ limit of the XXZ chain is considered, leading to integrable dynamics~\cite{Zadnik2021a,Zadnik2021b}.
The key difference in our work, compared to the previous literature, consists of having a chiral kinetic term, whereas in the mentioned works symmetric constraints are considered.

\begin{figure}[t]
\centering
\includegraphics[width=.65\columnwidth]{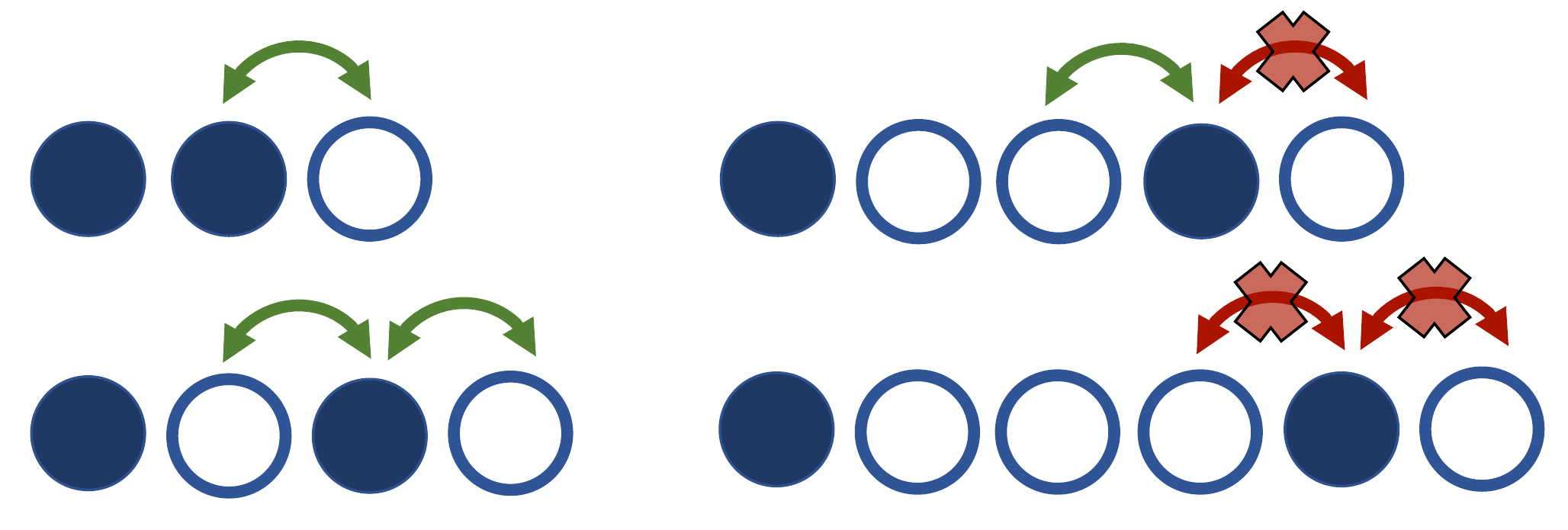}
\caption{\label{Fig:cartoon}
Illustration of constrained hopping in the range-2 particle conserving East model. }
\end{figure}

Hamiltonians $\hat H_r$ for all values of $r$ feature $U(1)$ symmetry related to the conservation of total boson number, justifying the name of particle-conserving East models. In this work we mostly focus on the case of $r=2$ with homogeneous hopping parameters $t_\ell=1$, as written in Eq.~(\ref{Eq:Hr=2}). 
We discuss the generality of our results with respect to the choice of hopping strengths and range of constraint in Appendices~\ref{App:generic r} and ~\ref{App:generic dyn}.
A major feature of this family of models is Hilbert space fragmentation, which is known to affect spectral and dynamical properties.
As such we begin our investigation by looking into the nature of Hilbert space fragmentation in these models in Section~\ref{Sec:Eigenstates}, where we highlight the generality of our results, by formulating them for a general range $r$ and show examples for $r=2$. 

\section{Hilbert space fragmentation and eigenstates}\label{Sec:Eigenstates}%

In this Section we focus on the phenomenon of Hilbert space fragmentation in the particle-conserving East models introduced above. First, we discuss the block structure of the Hamiltonian in the product state basis --- known as a classical Hilbert space fragmentation --- and define the largest connected component of the Hilbert space. Next, in Sec.~\ref{Sec:Quantum Fragmentation} we discuss the emerging disconnected components of the Hilbert space that are not manifest in the product state basis, leading to quantum Hilbert space fragmentation. 

\subsection{Classical Hilbert space fragmentation} \label{Sec:Classical fragmentation}

Due to the $U(1)$ symmetry of the Hamiltonian~(\ref{Eq:Hr}), the global Hilbert space is divided in blocks labeled by the different number of bosons $N_p$ with dimension given by the binomial coefficient $\mathcal{C}^L_{N_p}$. Within each given sector of total particle number $N_p$, the constrained hopping causes further fragmentation of the Hilbert space in extensively many subspaces.  First, the leftmost boson in the system is always frozen. Hence, as we discussed in Section~\ref{Sec:Model}, we choose the first site to be always occupied, which may be viewed as a boundary condition. In addition, a boson may also be frozen if the number of particles to its left is too small. An example configuration is given by the product state $\ket{\bullet\circ\circ\circ\bullet\bullet\circ\circ}$ for the $r=2$ model, where $\circ$ corresponds to an empty site and $\bullet$ is a site occupied by one boson. Here the second boson cannot move since the previous two sites are empty and cannot be occupied. 

In view of this additional fragmentation, we focus on the largest classically connected sector of the Hilbert space with a fixed number of particles, $N_p$. This sector can be constructed starting from a particular initial state $\ket{\text{DW}}$, where all particles are located at the left boundary,
\begin{equation}
\label{Eq:psi0}
\ket{\text{\text{DW}}} = |\underbrace{\bullet\bullet\bullet\dots\bullet}_{N_p}\underbrace{\circ\circ\circ\dots\circ}_{L-N_p}\rangle.
\end{equation}
Starting from this initial state the constraint will limit the spreading of particles, that can reach at most 
\begin{equation}\label{Eq:Lstar}
L^*_r(N_p)=(r+1)N_p-r
\end{equation} 
sites, corresponding to the most diluted state, $\ket{\bullet\circ\circ\bullet\circ\circ\bullet\circ\circ\bullet\ldots}$ for $r=2$. Thus, in what follows we use the system size $L=L^*_r$ uniquely defined by the number of particles and the range of the constraint in Eq.~(\ref{Eq:Lstar}).

The fragmentation of the Hilbert space discussed above may be attributed to a set of emergent conserved quantities in the model in addition to the total particle number, $\hat N_\text{tot} = \sum_i \hat n_i$. The first class of conserved operators responsible for the freezing of the leftmost particle is written as
\begin{equation}
\label{Eq:Op fisrt site}
\hat{N}_{\ell_0} = \ell_0 \bigr[\prod_{i<\ell_0} (1-\hat{n}_i)\bigr] \hat{n}_{\ell_0}.
\end{equation}
Since projectors in this operator are complementary to the projectors in the Hamiltonian, this satisfies the property $\hat{N}_{\ell_0} \hat H_r = \hat H_r \hat{N}_{\ell_0}  = 0$, hence trivially having a zero commutator. This conservation law induces further fragmentation of the Hilbert space into $L-N_p$ sectors labeled by the position of the leftmost boson. 

The second class of operators yields a further fragmentation within each sector with fixed position of the leftmost particle. One can check that a region of $N_\text{left}$ particles with length $L_{r}^{\text{left}} \geq L^*_r(N_\text{left}) +r+1$ is dynamically isolated from any configuration on its right. By construction, the particles in the left part can never facilitate hopping of particles on the right, as they always have a distance $d>r$, hence different sectors can be labeled by the position and width of the frozen regions. The simplest example of such configuration is given by $\ket{\bullet\circ\circ\bullet\circ\circ\circ\bullet\dots}$ for $r=2$. Formally, these conserved quantities are represented by the following operator
\begin{equation}
\label{Eq:Op intermediate dark states}
\hat{O}^{L_{r}^{\text{left}}}_{N_\text{left}} =\hat{\mathcal{P}}_{N_\text{left}} \Bigr[\prod_{k=L_r^*(N_\text{left})+1}^{L_{r}^{\text{left}} - L^*(N_\text{left})} (1-\hat{n}_k)\Bigr] \hat{n}_{L_{r}^{\text{left}}+1},
\end{equation}
where $\hat{\mathcal{P}}_{N_\text{left}}$ is the projector on the states with $N_\text{left}$ particles in the first $L^*_r(N_\text{left})$ sites. The freedom in the choice of $L_{r}^{\text{left}}$ yields $r(N_p-N_\text{left}-1)$ different sectors for a fixed $N_\text{left}$. Hence, the number of the fragmented sectors is given by
\begin{equation}
\label{Eq:fragm frozen regions}
\sum_{N_\text{left}=1}^{N_p-1} r(N_p-N_\text{left}-1) = \bigr[\frac{1}{2}(N_p^2-3N_p)+1\bigr] \propto N_p^2.
\end{equation}
We notice that additional levels of fragmentation can emerge whenever the right part can be further decomposed in a similar way to the one discussed above. Every time that happens, additional subsectors appear for some of the sectors identified by the operator $\hat{O}^{L_{r}^{\text{left}}}_{N_\text{left}}$. As the number of additional levels of fragmentation increases proportionally to $N_p$, each adding subsectors to the previous level, one finally obtains that the asymptotic behavior of the global number of classically fragmented subsectors has to be $O(\exp(N_p))$. The exponential increase of the number of disconnected subsectors was verified numerically, thus properly identifying a case of Hilbert space fragmentation.

\begin{figure}[t]
\centering
\includegraphics[width=.75\columnwidth]{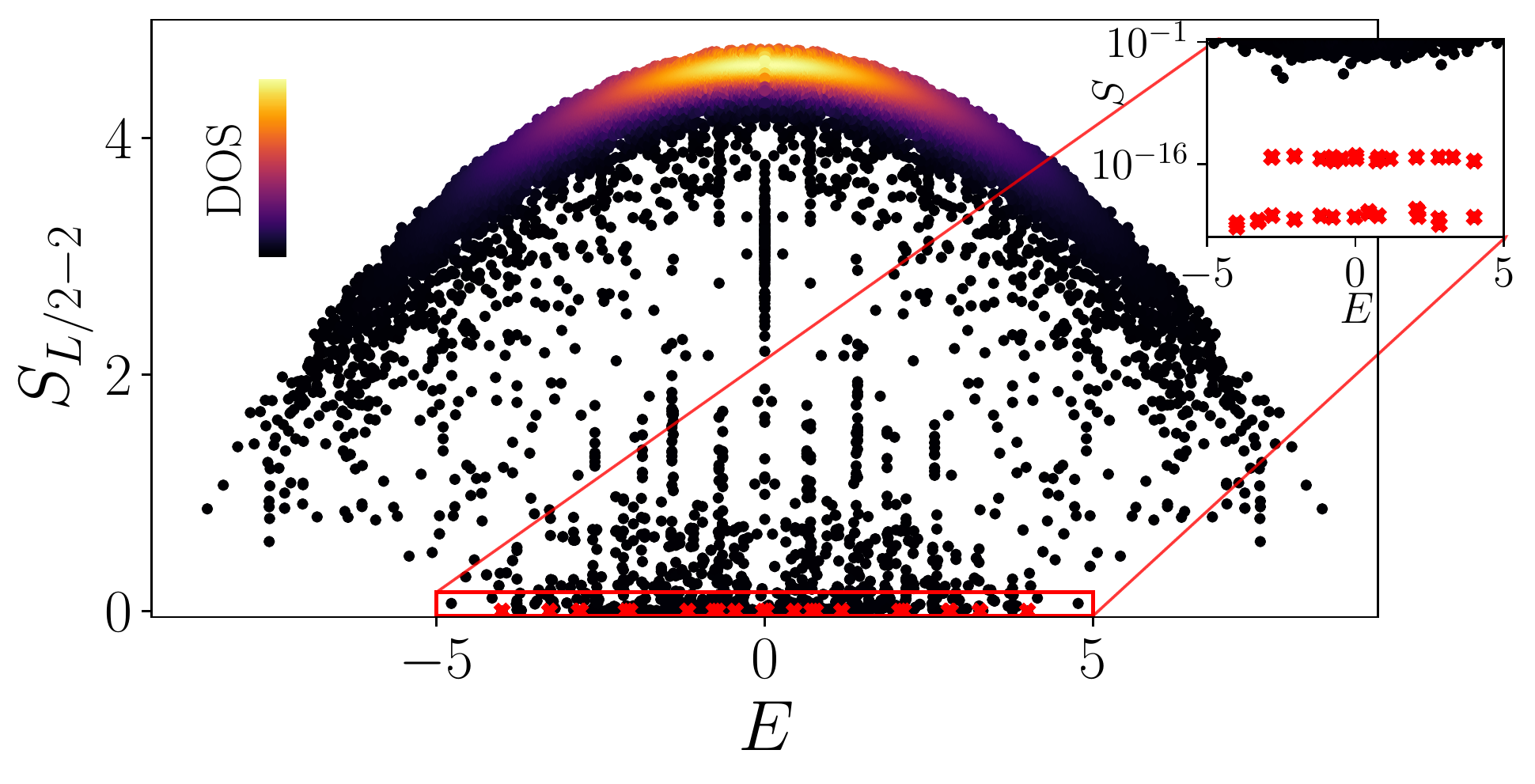}
\caption{\label{Fig:S evecs}
Entanglement entropy of eigenstates along the bipartite cut at the site $8$ for $N_p=8$ and $L=22$. The color intensity corresponds to the density of dots, revealing that the majority of the eigenstates have nearly thermal entanglement. However, a large number of eigenstates has entanglement much lower than the thermal value. Among these, the red dots correspond to entanglement being zero up to numerical precision (inset). 
}
\end{figure}

\subsection{Recursive quantum Hilbert space fragmentation}\label{Sec:Quantum Fragmentation}%

Due to the fragmentation of the Hilbert space in the computational basis discussed above, we focus on the largest sector of the Hilbert space as defined in the previous section. In Appendix~\ref{App:therm} we show that the statistic of the level spacing for the Hamiltonian $\hat H_2$ within this block follows the Wigner-Dyson surmise, confirming that we resolved all symmetries of this model and na\"ively suggesting an overall thermalizing (chaotic) character of eigenstates~\cite{D'Alessio2016}.

To further check the character of eigenstates, we consider their entanglement entropy. We divide the system into two parts, $A$ containing sites  $1,\ldots, i$,  $A=[1,i]$ and its complement denoted as $B=[i+1,L]$. The entanglement entropy of the eigenstate $\ket{E_\alpha}$ for such bipartition is obtained as the von Neumann entropy of the reduced density matrix $\rho_i = \tr_B \ket{E_\alpha}\bra{E_\alpha}$
\begin{equation}
\label{Eq:Entanglement}
S_i = -\tr\bigr[\rho_i\ln\rho_i\bigr].
\end{equation}
In thermal systems entanglement of highly excited eigenstates is expected to follow volume law scaling, increasing linearly with $i$ for $i\ll L$, and reaching maximal value for $i=L/2$. However, our numerical study of the entanglement entropy shows strong deviations from these expectations, in particular revealing a significant number of eigenstates with extremely low, and even exactly zero, entanglement, a feature typical of quantum many-body scars~\cite{Turner2018,Michailidis2018,Bernevig2018,Choi2019,Iadecola2019,Motrunich2019,Papic2020,Knolle2020,Serbyn2021,Regnault2022}.

Figure~\ref{Fig:S evecs} illustrates such anomalous behavior of eigenstate entanglement for a chain of $L=22$ sites. For the bipartite cut shown, $A=[1,8]$, most of the eigenstates have increasing entanglement as their energy approaches zero, where the density of states is maximal, in agreement with thermalization. Nevertheless, a significant number of eigenstates features much lower values of entanglement, and the red box and inset in Fig.~\ref{Fig:S evecs} highlight the presence of eigenstates with zero entanglement (up to numerical precision). We explain this as a result of an additional fragmentation of the Hilbert space caused by the interplay of the constraint and boson number conservation. 

Eigenstates with zero entanglement, denoted as $\ket{E_{S=0}}$, are separable and can be written as a product state of the wave function in the region $A$ and in its complement $B$. To this end, we choose the wave function $\ket{\psi^\ell_m}$ of the separable state $\ket{E_{S=0}}$ in the region $A$ as an eigenstate of the Hamiltonian $\hat H_r$ restricted to the Hilbert space of $m$ particles in $\ell$ sites. The state $\ket{\psi^\ell_m}$ has to satisfy the additional condition $\bra{\psi_m^\ell}\hat{n}_{\ell}\ket{\psi_m^\ell}=0$, i.e. that the last site of the region is empty. Provided such state exists, we construct the separable eigenstate $\ket{E_{S=0}}$ as 
\begin{equation}
\label{Eq:zero S evec}
\ket{E_{S=0}} = \ket{\psi_m^\ell}\otimes \underbrace{\ket{\circ\circ\dots \circ}}_{q}\otimes\ket{\psi_R},
\end{equation}
where  $\ket{\psi_R}$ is an eigenstate of the Hamiltonian restricted to $L-\ell-q$ sites and $N_p-m$ particles. Inserting an empty region of $q\geq r$ sites separating the support of $\ket{\psi_m^\ell}$ and $\ket{\psi_R}$ ensures that the two states are disconnected. 
Note that $q$ is upper bounded by the requirement that the resulting state belongs to the largest classically fragmented sector. 
It is easy to check that the state~$\ket{E_{S=0}}$ is an eigenstate of the full Hamiltonian. Indeed, thanks to the empty region $q$ the particles in $A$ cannot influence those in $B$ and the two eigenstates of the restricted Hamiltonian combine into an eigenstate of the full system.

The construction of  $\ket{E_{S=0}} $ relies on the existence of eigenstates $ \ket{\psi_m^\ell}$ with vanishing density on the last site. This is a non-trivial requirement that \emph{a priory} is not expected to be satisfied. However, we observe that such eigenstates can be found within the degenerate subspace of eigenstates with zero energy, see Appendix~\ref{App:PsiL}. If  $\ket{\psi_m^\ell}$ is an eigenstate with zero energy, the energy of eigenstate  $\ket{E_{S=0}}$ is determined only by the energy of the $\ket{\psi_R}$.
The existence of $ \ket{\psi_m^\ell}$ relies on two conditions which have to hold simultaneously: $\ell>m+r$ and $(r+1)m-r\ge\ell$. 
These are satisfied only for $m\geq 3$ particles, thus resulting in a minimal size of the left region $\ell_\text{min}=6$ for $r=2$.
While there is no guarantee that states $\ket{\psi_m^\ell}$ exist for generic $(m,\ell)$, we have an explicit analytic construction for the smallest state $\ket{\psi_3^6}$ for $(m,\ell)=(3,6)$
\begin{equation}\label{Eq:min-left}
\ket{\psi_3^6} = \frac{1}{\sqrt{2}}\bigr[\ket{\bullet\bullet\circ\circ\bullet\circ}-\ket{\bullet\circ\bullet\bullet\circ\circ}\bigr],
\end{equation}
similarly we report solutions up to $(m,\ell)=(7,18)$ in Appendix~\ref{App:PsiL}. 
Furthermore, for each $(m,\ell)$ satisfying the condition, one can easily verify that stacking multiple $|\psi_m^\ell\rangle$ separated by at least $r$ empty sites generates another state fulfilling the same condition. 
This recursive construction of the left states in Eq.~(\ref{Eq:zero S evec}), together with the explicit example Eq.~(\ref{Eq:min-left}), guarantees the existence of an infinite number of $\ket{\psi_m^\ell}$, in the thermodynamic limit. We further notice that a similar decomposition can be applied to the right eigesntates, $\ket{\psi_R}$ in a recursive fashion. The observed \textit{recursive} quantum Hilbert space fragmentation is a novel feature of this family of Hamiltonians~(\ref{Eq:Hr}), relying on both particle conservation and chiral constraint.

\begin{figure}[t]
\centering
\includegraphics[width=1.01\columnwidth]{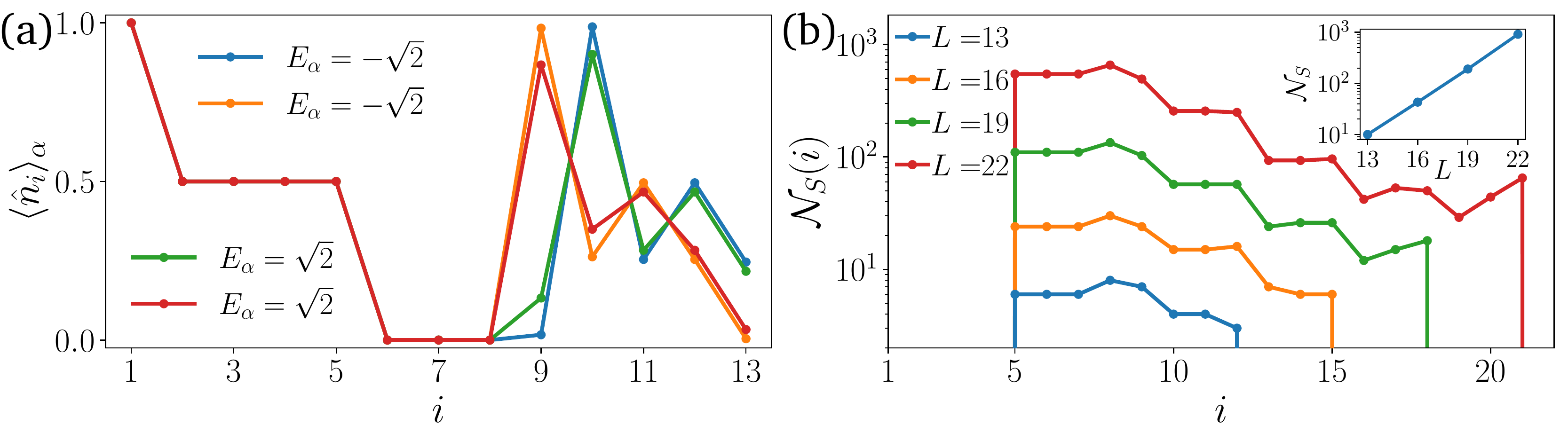}
\caption{\label{Fig:Special evecs}
(a): The density profile of the zero-entanglement eigenstates for $L=13$ shows a common pattern, due to their special structure~(\ref{Eq:zero S evec}). The first sites correspond to the zero mode of the Hamiltonian restricted to $3$ particles in $6$ sites $\ket{\psi_3^6}$, followed by $2$ empty sites. The right subregion can then be any of the $6$ eigenstate of $H$ for $2$ particles in $4$ sites, with energy $\pm\sqrt{2},\;0$. We notice that eigenstates with the same $\ket{\psi_R}$ but a different number of empty sites separating it from $\ket{\psi_m^\ell}$ are degenerate and can be mixed by the numerical eigensolver, as is the case in the density profiles shown here.
(b): The number of zero entanglement entropy eigenstates $\mathcal{N}_S(i)$ depends on the boundary of the subregion $A=[1,i]$. 
In particular, in the interval $i\in[5,9]$ the number of zero-entanglement eigenstates is exponentially larger compared to more extended left subregions. At larger $i$ recursively fragmented eigenstates contribute to $\mathcal{N}_S(i)$ for $L\geq13$.  The total number of zero-entanglement eigenstates, $\mathcal{N}_S$, grows exponentially in $L$, as shown in the inset. Note that $\mathcal{N}_S\neq\sum_i\mathcal{N}_S(i)$, as some eigenstates have zero entanglement across multiple bipartite cuts.
 }
\end{figure}

Let us explore the consequence of the existence of the special eigenstates defined in Eq.~(\ref{Eq:zero S evec}). Given the special character of the wave function $\ket{\psi_m^\ell}$, we expect that states $\ket{E_{S=0}} $ have a similar pattern of local observables in the first $\ell$ sites. An example of such behavior is shown in Figure~\ref{Fig:Special evecs}(a), which reveals that all four states $\ket{E_{S=0}}$ that have zero entanglement across at least one bipartite cut in the $L=13$ chain for $r=2$ feature the same density expectation values, $\langle \hat{n}_i\rangle_\alpha=\bra{E_\alpha}\hat{n}_i\ket{E_\alpha}$, in the first $\ell=6$ sites. Starting from the site number $i=9$, the density profile has different values on different eigenstates, corresponding to different wave functions $\ket{\psi_R}$ in Eq.~(\ref{Eq:zero S evec}).

The number of eigenstates with zero entanglement grows exponentially with system size.  Even for the case of a fixed $\ket{\psi_m^\ell}$, the right restricted eigenstate $\ket{\psi_R}$ is not subject to any additional constraints, hence the number of possible choices of $\ket{\psi_R}$ grows as the dimension of the Hilbert space of $N_p-m$ particles on $L-\ell-r$ sites, that is, at fixed $m$, asymptotically exponential in $N_p$. In the general case where $(m,\ell)$ are allowed to change, new $\ket{E_{S=0}}$ states will appear, with zero entanglement entropy at different bipartite cuts, according to the size of the left region. Finally, the recursive nature of the fragmentation discussed above is expected to give eigenstates with zero entropy across two or more distinct cuts which are separated by a non-vanishing entanglement region. These states are observed in numerical simulations starting from $N_p=7$ and $L=19$.

To illustrate the counting of eigenstates with zero entropy at a cut separating subregion $A=[1,i]$ from the rest of the system, we denote their number as $\mathcal{N}_S(i)$. For $i<5$, this number is zero $\mathcal{N}_S(i)=0$, as explained in the construction of these states. 
For $i\geq 5$ we observe a large $\mathcal{N}_S(i)$, exponentially increasing with system size.
However, at larger $i$, the available configurations that can support states of the form Eq.~(\ref{Eq:zero S evec}) decrease and $\mathcal{N}_S(i)$ drops and eventually vanishes. As $N_p$ and system size increase, left states $\ket{\psi_m^\ell}$ with a larger support $\ell$ are allowed thus increasing the range of sites where $\mathcal{N}_S(i)>0$. This is also due to recursive fragmentation which can appear starting from $N_p=5$ and $L=13$. Carefully counting all \textit{distinct} eigenstates $\ket{E_{S=0}}$ we confirm that their total number $\mathcal{N}_S$  grows exponentially with system size in the inset of Fig.~\ref{Fig:Special evecs}(b)

\section{Dynamics}\label{Sec:Dyn}

After discussing recursive quantum Hilbert space fragmentation in the particle-conserving East model, we proceed with the study of the dynamics. First, in Section~\ref{Sec:QDyn} we consider the dynamical signatures of Hilbert space fragmentation. Afterwards, in Section~\ref{Sec:Qdyn DW} we discuss the phenomenology of particle spreading starting from a domain wall state and illustrate how this can be connected to the structure of the Hilbert space. Finally, we compare the quantum dynamics to that of a classical cellular automaton in Section~\ref{Sec:ClDyn}.

\subsection{Dynamical signatures of quantum Hilbert space fragmentation}\label{Sec:QDyn}

The zero-entanglement eigenstates $\ket{E_{S=0}}$ identified in Eq.~(\ref{Eq:zero S evec}) span a subsector of the Hilbert space which is dynamically disconnected from the rest. In this subspace the Hamiltonian has non-trivial action only in the right component of the state, and eigenstates can be written as product states across the particular cut. Below we discuss signatures of such fragmentation in dynamics launched from weakly entangled initial states. 

\begin{figure}[h]
\centering
\includegraphics[width=.75\linewidth]{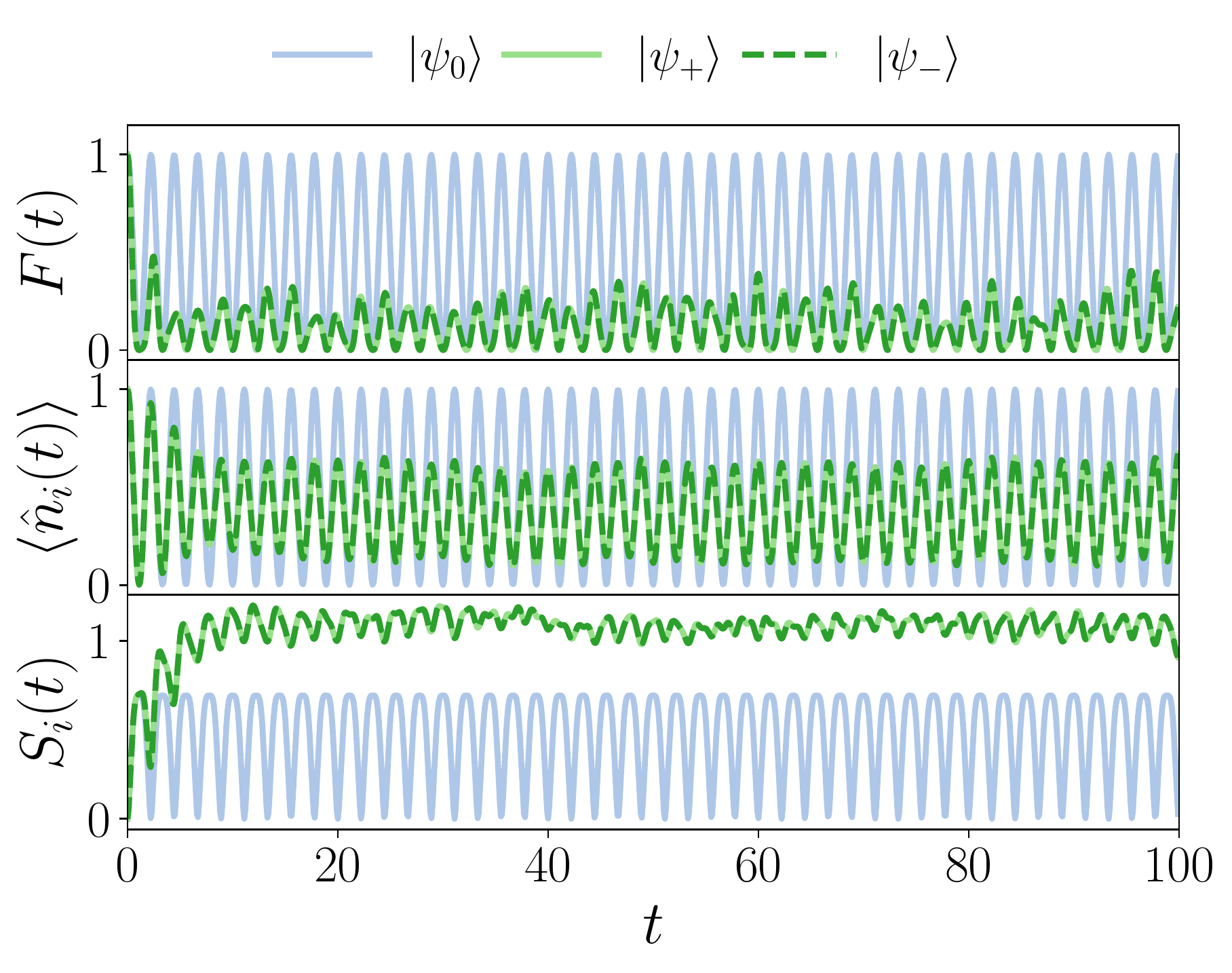}
\caption{\label{Fig:revival}
The signatures of quantum Hilbert space fragmentation can be observed for initial states that have a large overlap with zero-entanglement eigenstates $\ket{E_{S=0}}$. 
The fidelity $F(t)=|\bra{\psi_0}\psi(t)\rangle|^2$ shows periodic revivals for all three initial states; choosing an eigenstate on the left portion of the chain results in perfect revivals (blue curve). Entanglement entropy across the cut $i=11$ in the middle of the right region $R$ and density on the same site show oscillations with identical frequency.
}
\end{figure}

As an illustrative example, we show in Figure~\ref{Fig:revival} the time evolution of a state of the form defined in Eq.~(\ref{Eq:zero S evec}) for $L=13$. To obtain non-trivial dynamics, we replace the eigenstate $\ket{\psi_R}$ with a product state. In particular, we choose the initial state as
\begin{equation}
\label{Eq:psi0 revivals}
\ket{\psi_0} = \frac{\ket{\bullet\bullet\circ\circ\bullet\circ}-\ket{\bullet\circ\bullet\bullet\circ\circ}}{\sqrt{2}}\otimes\ket{\circ\circ}\otimes\ket{\bullet\circ\bullet\circ\circ},
\end{equation}
and consider the time-evolved state $\ket{\psi_0(t)} =e^{-\imath t \hat H_2} \ket{\psi_0}$.
The action of the full Hamiltonian does not affect the left part of the state and the Hamiltonian acting on the last five sites in the chain $R=[9,13]$ is a simple $3\times 3$ matrix
\begin{equation}
\label{Eq:HR}
\hat{H}_R = \begin{pmatrix}
0 &1 &0\\
1 &0 &1\\
0 &1 &0
\end{pmatrix}.
\end{equation}
in the $\{\ket{\bullet\bullet\circ\circ\circ},\ket{\bullet\circ\bullet\circ\circ},\ket{\bullet\circ\circ\bullet\circ}\}$ basis. Diagonalizing this matrix, we write the time-evolved state $\ket{\psi_0(t)}$ as
\begin{equation}
\label{Eq:psi t Rev}
\begin{split}
\ket{\psi(t)} =& \ket{\psi_m^\ell}\otimes\ket{00}\otimes\Big[\cos(\sqrt{2}t)\ket{\bullet\circ\bullet\circ\circ}\\
&
 - \sin(\sqrt{2}t)\frac{\ket{\bullet\bullet\circ\circ\circ}+\ket{\bullet\circ\circ\bullet\circ}}{\sqrt{2}}\Big],
\end{split}
\end{equation}
hence the fidelity reads $F(t) = |\bra{\psi_0}\psi(t)\rangle|^2 = \cos^2(\sqrt{2}t)$. As the time-evolution in Eq.~(\ref{Eq:psi t Rev}) involves only three different product states, it produces perfect revivals with period $T=\pi/\sqrt{2}$. This periodicity also affects observables, such as the density in the region $R$, and the entanglement entropy.

This periodic dynamics also appears in the two product states $\ket{\psi_+}=\ket{\bullet\bullet\circ\circ\bullet\circ\circ\circ\bullet\circ\bullet\circ\circ}$ and $\ket{\psi_-}=\ket{\bullet\circ\bullet\bullet\circ\circ\circ\circ\bullet\circ\bullet\circ\circ}$ that are contained in Eq.~(\ref{Eq:psi0 revivals}). These states indeed show revivals of the fidelity with the same period $T$, although the peaks are more suppressed. This is not surprising, as these states have only part of their weight in the disconnected subspace.

In Figure~\ref{Fig:revival} we show the results of the dynamics of the state $\ket{\psi_0}$, Eq.~(\ref{Eq:psi0 revivals}), together with the two product states generating the superposition, $\ket{\psi_\pm}$. In addition to fidelity, we also show the density and entanglement dynamics of sites $i$ within the right region $R$.
As expected, the fidelity shows revivals with period $T=\pi/\sqrt{2}$, and similar oscillations are also observed in local operators and entanglement. 
\begin{figure}[b!]	
\centering
\includegraphics[width=.65\columnwidth]{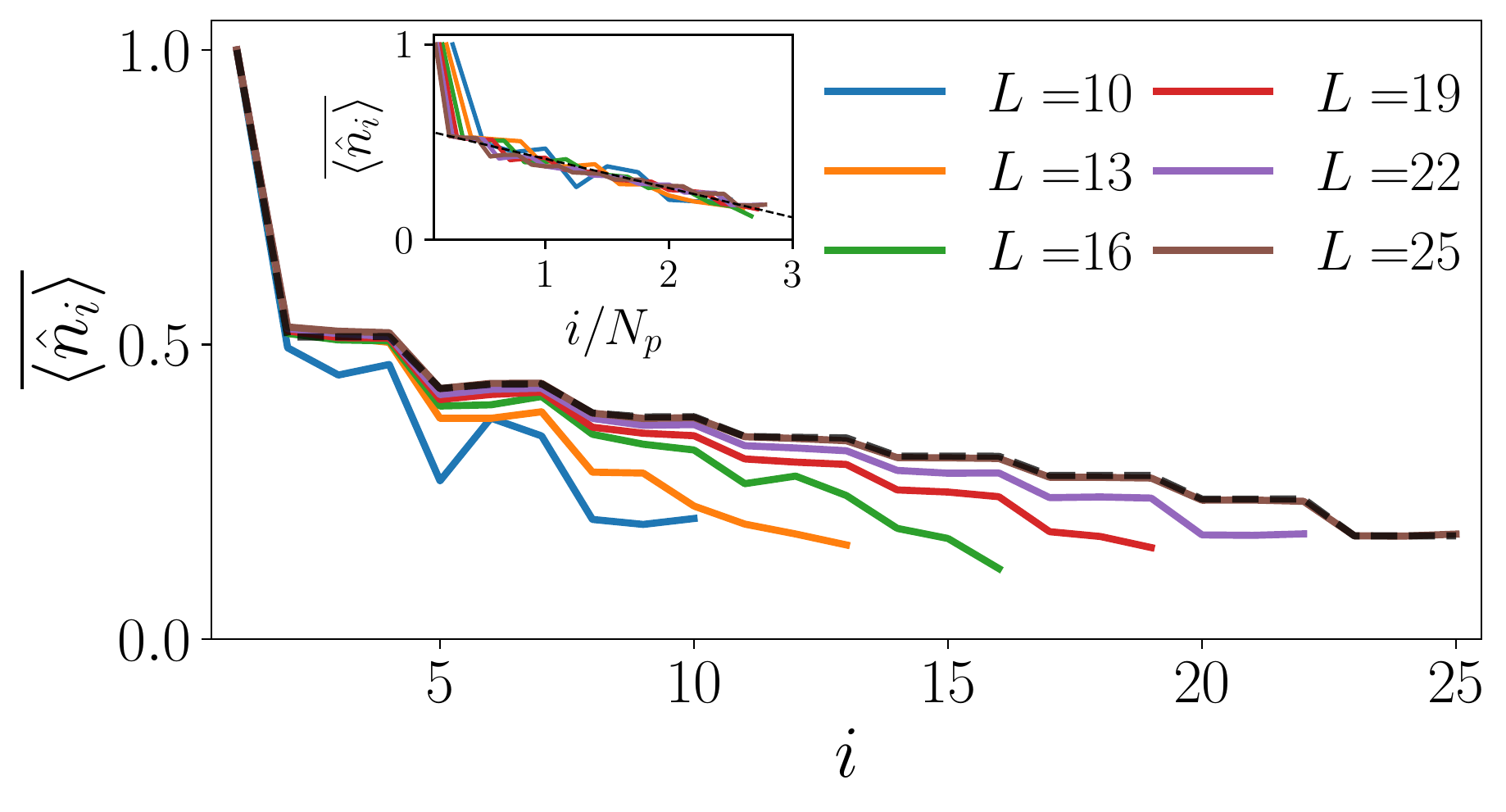}
\caption{\label{Fig:n saturation}
The constrained character of the model leads to a non-uniform stationary density profile for the domain wall initial state. This coincides with the infinite-temperature prediction on large systems, as highlighted by the dashed line corresponding to $\tr[\hat{n}_i]/\tr[\mathbb{1}]$ for $L=25$, where $\tr[\hat{O}]=\sum_j \hat{O}_{jj}$. Rescaling the $x$-axis by the number of particles $N_p$, we obtain a good collapse of the data, as shown in the inset. The particle density follows a linear decrease $\overline{\langle \hat{n}_i\rangle}\approx \overline{\langle \hat{n}_2\rangle}-c(i-2)/N_p$, with $c\approx0.15$.
}
\end{figure}While the initial state $\ket{\psi_0}$ defined in Eq.~(\ref{Eq:psi0 revivals}) presents perfect revivals with $F(T)=1$, the  product states $\ket{\psi_\pm}$ does not display perfect fidelity revivals show larger entanglement. We note, that since the two product states  $\ket{\psi_\pm}$ together form a state $\ket{\psi_m^\ell}$ their dynamics in the region $R$ is not affected by the choice of the left configuration, and all considered quantities for theses two initial states have identical dynamics.

\subsection{Phenomenology of dynamics from the $\ket{\text{\text{DW}}}$ initial state}\label{Sec:Qdyn DW}

\begin{figure}[b!]
\includegraphics[width=1.01\textwidth]{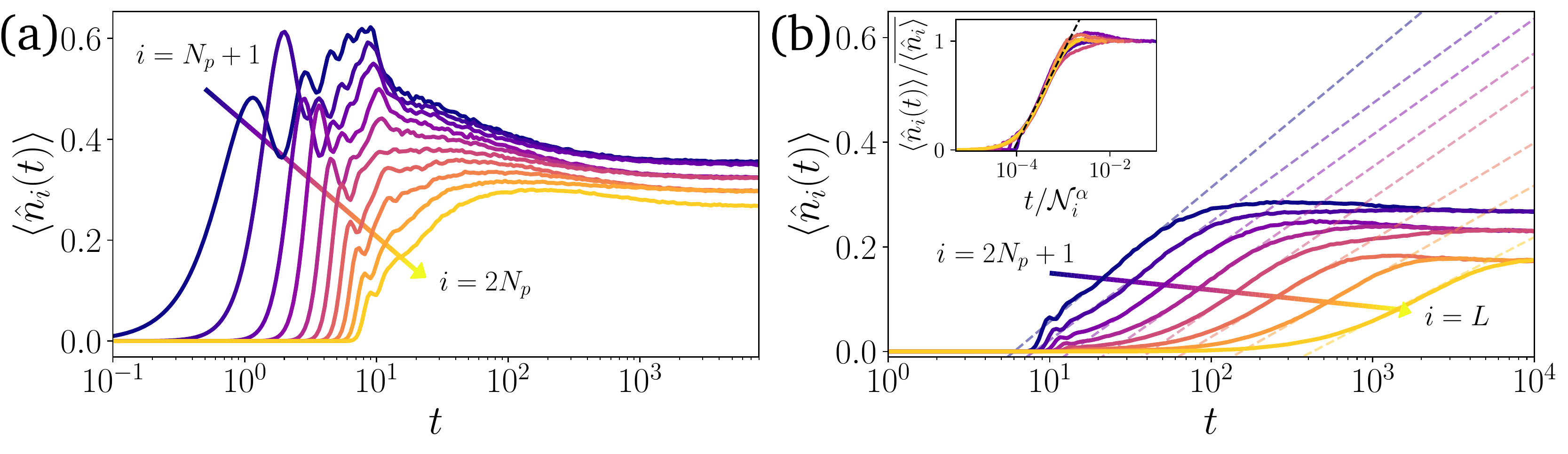}
\caption{\label{Fig:n dynamics L28}
The approach to saturation in the density dynamics is very different depending on the region within the chain. (a) In the first $2N_p$ sites of the chain a fast relaxation takes place due to the weak role of the constraint in dense regions. (b) For the right part of the chain, $i>2N_p$ anomalously slow logarithmic dynamics arise. The inset shows the data collapse upon rescaling the density axis by the long time average and the time axis by the number of states within each \textit{leg} of the graph $\mathcal{N}_i$ to the power $\alpha\approx 1.15$, as discussed in more detail at the end of this section. The data shown here are for a system of $L=28$ sites with $N_p=10$ bosons.
}
\end{figure}

After exploring the dynamics resulting from quantum Hilbert space fragmentation, we now turn to the dynamics in the remainder of the constrained Hilbert space focusing on the domain wall state~(\ref{Eq:psi0}). The domain wall state does not have any overlap with zero entanglement eigenstates except for possibly states with zero entanglement on the last cut. It is also characterized by a vanishing expectation value of the Hamiltonian, corresponding to zero energy density, where the density of states is maximal. Hence, thermalization implies that time evolution from the domain wall leads to the steady state where all observables agree with their infinite-temperature expectation value. To check this property we focus on the expectation value of the particle density operators throughout the chain.

Figure~\ref{Fig:n saturation} shows the infinite time average of the particle density, $\overline{\langle \hat{n}_i\rangle}$ obtained through the diagonal ensemble 
\begin{equation}
\overline{\langle \hat{n}_{i}\rangle}=\sum_\alpha |\bra{\text{DW}}E_\alpha\rangle |^2 \bra{E_\alpha}\hat{n}_i\ket{E_\alpha},
\end{equation}
where the sum runs over all eigenstates $\alpha$. This calculation is performed for $L\leq 22$, where the full set of eigenstates can be obtained through exact diagonalization. For larger systems, the infinite time average value of $\overline{\langle \hat{n}_i\rangle}$ is approximated as the average of the density in the time-window $t\in [6.9\times10^3,10^4]$. We observe that the density profile agrees well with the infinite-temperature prediction. See Appendix~\ref{App:therm} for details of the calculation. 

The infinite-temperature prediction for the density profile does not result in a homogeneous density due to the constraint. The number of allowed configurations with non-zero density in the last sites is indeed limited by the constraint, and results in a lower density in the rightmost parts of the chain.
In addition, the profile has a step-like shape that is related to the range-2 constraint in the model.  In the inset of Fig.~\ref{Fig:n saturation} we show that the density profiles collapse onto each other when plotted as a function of $i/N_p$. This suggests the heuristic expression for the density profile $\overline{\langle \hat{n}_i\rangle}\approx \overline{\langle \hat{n}_2\rangle}-c (i-2)/{N_p}$ where $c\approx0.15$ is a positive constant. 

Although the saturation profile of the density is consistent with thermalization, below we demonstrate that \emph{relaxation} to the steady state density profile is anomalous. The time-evolution of the density $\langle \hat{n}_i(t)\rangle = \bra{\psi(t)}\hat{n}_i\ket{\psi(t)}$ is shown in Figure~\ref{Fig:n dynamics L28} for $L=28$ sites up to times $t\approx 10^4$. The data demonstrates that the relaxation of density qualitatively depends on the location within the chain. In the left part of the chain with $i\leq 2N_p$, the spreading of the density front is fast, and saturation is reached quickly on timescales of $O(10)$, as shown in Fig.~\ref{Fig:n dynamics L28}(a). This can be attributed to the fact that the constraint is not effective at large densities. In contrast, in the rightmost part of the chain, $i>2N_p$ the constraint dramatically affects the spreading of particles resulting in the logarithmically slow dynamics in Fig.~\ref{Fig:n dynamics L28}(b). 

To further characterize the anomalous dynamics, we study the transport of the particle density on short time-scales for larger systems up to $L=37$ sites. For the systems with $L>28$ we use a fourth-order Runge-Kutta algorithm with a time-step as small as $\delta t=10^{-3}$. This allows us to reliably study the short-time behavior with sufficient accuracy down to $\delta t^4=10^{-12}$. We consider the dynamics of the root mean squared displacement, that measures the spreading of the \textit{center of mass}
\begin{equation}
\label{Eq:R tilde}
{R}(t) = \sqrt{\sum_{i>N_p} \langle\hat{n}_i(t)\rangle |i-N_p|^2}. 
\end{equation}

 \begin{figure}[t]
\centering
\includegraphics[width=1.01\columnwidth]{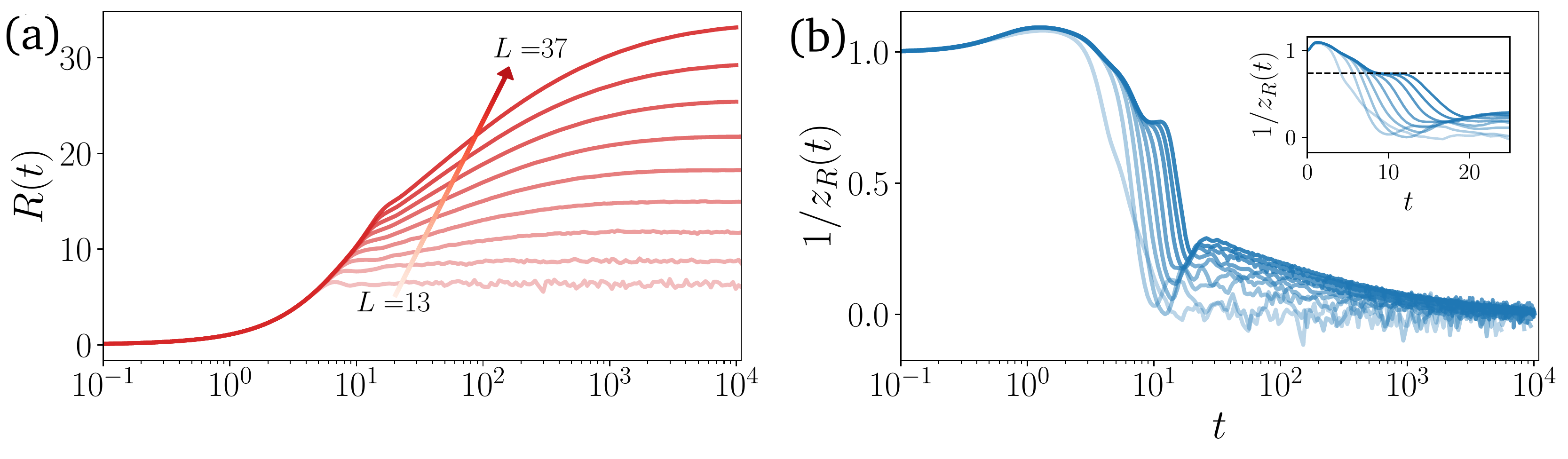}
\caption{\label{Fig:Rtilde and z}
(a) The behavior of the root-mean-square displacement shows an initial power-law growth $R(t)\sim t^{1/z_{{R}}(t)}$ followed by a slow-down to logarithmic behavior at later times, in agreement with the density dynamics. (b) The analysis of the dynamical exponent $z_{{R}}(t)$ shows the presence of a super-diffusive plateau $1/{z_{{R}}}\approx 0.74$ at intermediate times, whose duration grows with system size. At later times, the onset of logarithmic dynamics is signalled by the decay of $1/z_{{R}}(t)$. Data are for $13\leq L \leq 37$ from more to less transparent.
}
\end{figure}

The dynamics of $R(t)$ in Figure~\ref{Fig:Rtilde and z}(a) shows a clear initial power-law behavior drifting to much slower logarithmic growth at later times, in agreement with the dynamics of $\langle\hat{n}_i(t)\rangle$ in the right part of the chain. At even longer times $R(t)$ saturates to a value proportional to the system size $L$. Figure~\ref{Fig:Rtilde and z}(b) shows the instantaneous  dynamical exponent 
\begin{equation}\label{Eq:z-define}
z_{{R}}(t) =\left( \frac{d\ln R(t)}{d\ln t}\right)^{-1}.
\end{equation}
The early time dynamics is characterized by ballistic behavior, $z_R(t)\approx 1$ due to the large density in the vicinity of $i=N_p$. On intermediate time-scales $t\approx 10$,  a superdiffusive plateau of $1/z_{{R}}(t)\approx 0.74$ is visible. Finally, at longer times dynamics slow down and become logarithmic, consistent with a vanishing $1/z_R(t)$. Zooming in the time-window $t\leq30$, we notice that the extent of the superdiffusive plateau increases roughly linearly with system size, suggesting a persistence of the super-diffusive regime in the thermodynamic limit. 
The exact scaling, obtained by collapsing the curves in Fig.~\ref{Fig:Rtilde and z}(b), yields a power law behavior of the plateau extent,$t\sim L^{1.1}$, however due to the relatively small number of system sizes we cannot exclude a more natural $t\sim L$ dependence.

\begin{figure}[tb]
\centering
\includegraphics[width=.85\textwidth]{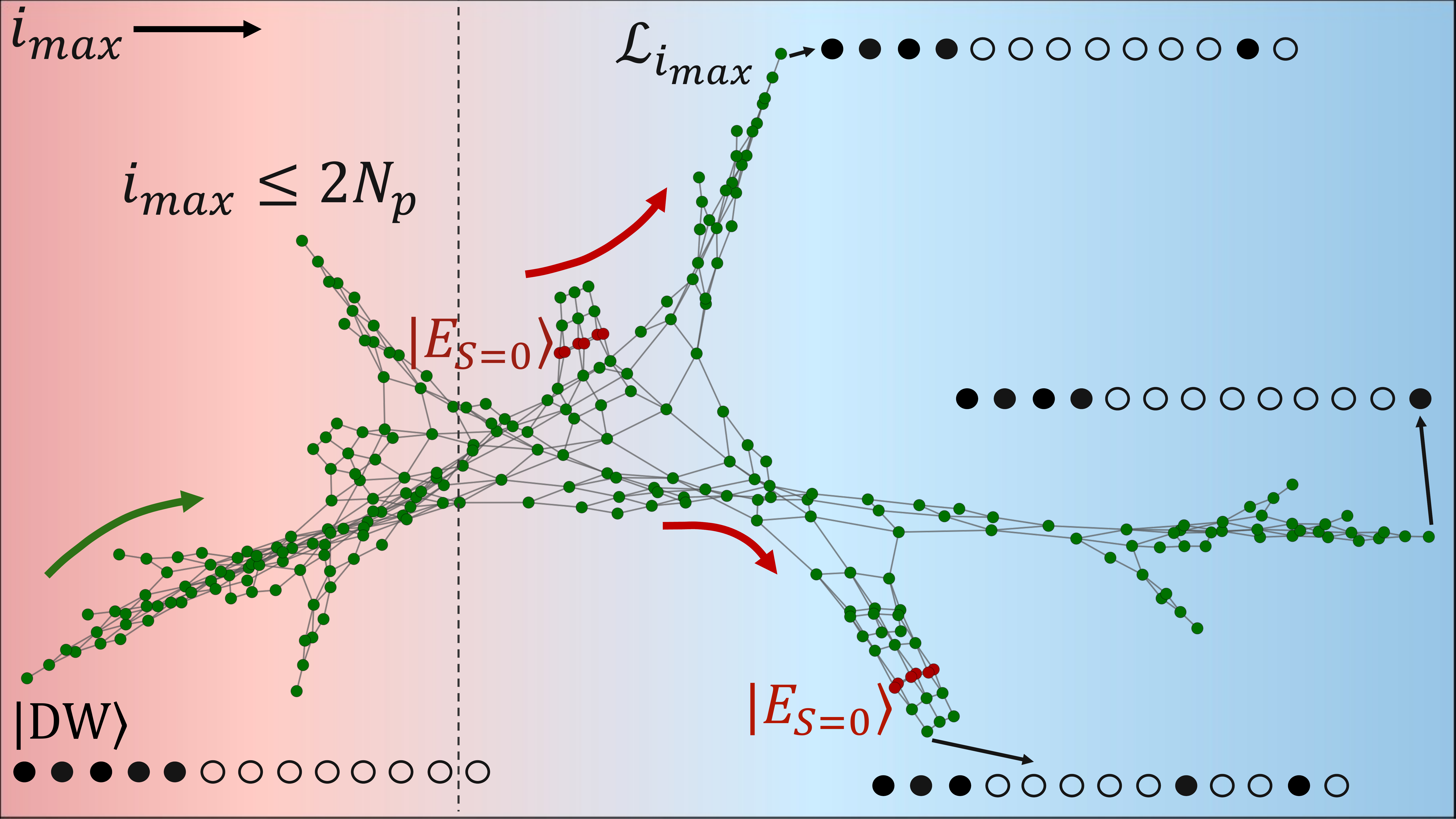}
\caption{\label{Fig:G} 
The representation of $\hat H_2$ as the adjacency graph $\mathcal{G}_r$ for system with $N_p=5$ particle and $L=13$ lattice sites. The dense central part -- \textit{backbone} -- has gradually decreasing number of vertices and connectivity as the position of the rightmost particle increases above $\imax>2N_p=10$ (dashed line). The \textit{legs} of the graph emanate from the backbone and correspond to regions where  $\imax$ is conserved.  The legs end with the product states (an example is labeled as $\mathcal{L}_{\imax}$), where a particular particle is frozen near the end of the chain.  Red vertices show product states corresponding to zero-entanglement eigenstates $\ket{E_{S=0}}$, which in this case have weight on $12$ out of $\mathcal{D}_{N_p}=273$ product states contained in the constrained Hilbert space.
}
\end{figure}

We now focus on capturing the phenomenology of the dynamics observed above using the structure of the Hamiltonian. In order to do so we interpret the Hamiltonian as a graph where the vertices of the graph enumerate the product states contained in a given connected sector of the Hilbert space. The edges of the graph connect product states that are related by any particle hopping process allowed by the constraint.  A particular example of such a graph for the system with $N_p=5$ particles and $L=13$ sites is shown in Fig.~\ref{Fig:G}.

The vertices of the graph in Fig.~\ref{Fig:G} are approximately ordered by the position of the rightmost occupied site $i_\text{max}\geq N_p$, revealing the particular structure emergent due to the constraint. The dense region that follows the domain wall product state has high connectivity, and we refer to it as the \textit{backbone}. In addition to the backbone, the graph has prominent \emph{legs} emanating perpendicularly. The legs are characterized by the conserved position of the rightmost particle that is effectively frozen due to the particles on the left retracting away, as pictorially shown in Fig.~\ref{Fig:G}.  Since such legs are in one-to-one correspondence with the position of the rightmost particle, $i_\text{max}$, their number grows linearly with system size. The number of product state configurations contained within each leg strongly depends on $i_\text{max}$. Given that the position of the rightmost particle is frozen within a leg, they cast a strong effect on the dynamics of the model.

In particular, the spreading of particles towards the right probed by $R(t)$ can be related to the presence of an increasing number of configurations within legs at large $\imax$, $\mathcal{N}_{\imax}$. These are characterized by long empty regions as the one depicted in Figure~\ref{Fig:G}, which require the collective motion of many particles to allow the hopping of the rightmost boson sitting at $\imax$. The slow dynamics observed, then, can be qualitatively understood as the effect of many states not contributing to the spreading and of the increasingly long empty regions that have to be crossed to activate hopping further to the right.
Looking back at the dynamics shown in Figure~\ref{Fig:n dynamics L28}, we highlight this effect by rescaling the time-axis by the number of configurations belonging to each leg, $\mathcal{N}_i$. The resulting collapse is shown in the inset of Figure~\ref{Fig:n dynamics L28}(b).

\subsection{Dynamics in constrained classical cellular automata}\label{Sec:ClDyn}

The anomalous relaxation of the quantum model from the domain wall state reported in Section~\ref{Sec:Qdyn DW} invites natural questions about the universality of dynamics in presence of inversion-breaking constraints. To shed light on this question, we introduce a classical cellular automaton model that  replaces the unitary time-evolution of the quantum model $\hat{U}(t) = \exp(-\imath\hat{H}t)$ with a circuit of local unitary gates preserving the same symmetries and constraints of the Hamiltonian~\cite{Nandkishore2019b,Pozsgay2021a}. 

\begin{figure}[tb]
\centering
\includegraphics[width=1.\columnwidth]{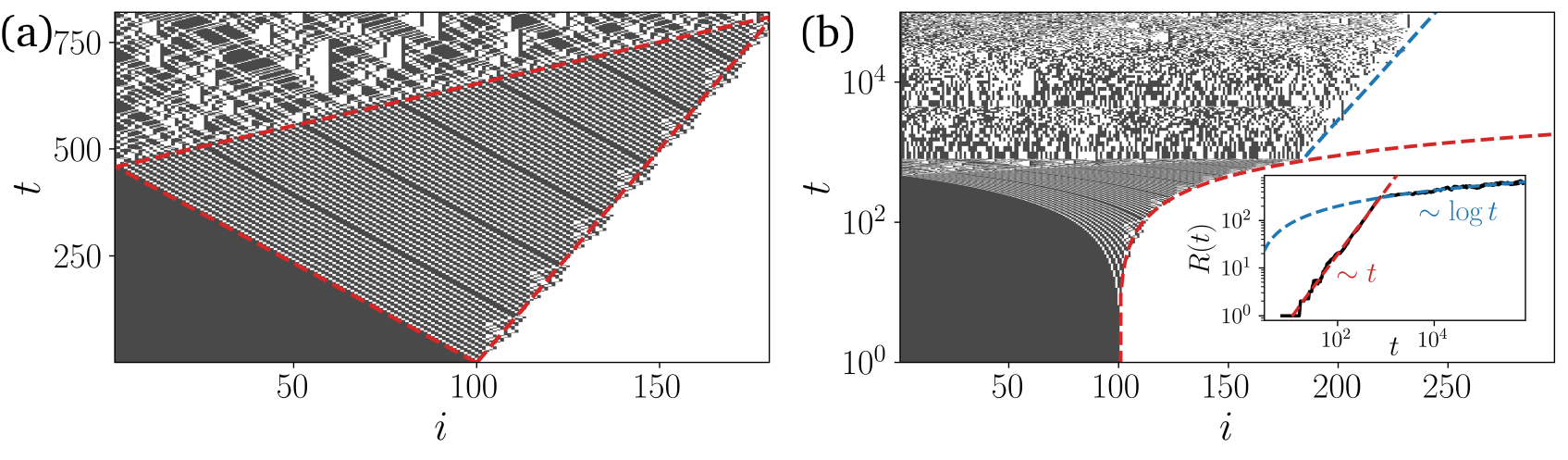}
\caption{\label{Fig:Circuit}
(a)-(b): Density evolution of the classical cellular automaton starting from domain wall initial state for a system with $L=298$ sites and $N_P=100$. (Black and white dots correspond to occupied and empty sites). (a) At short times particles spread ballistically into the empty region. Scattering events appear at regular time intervals at the boundaries of the red dashed triangle which defines the region of ballistic behavior. (b) At later times when particle density is lower the constraint becomes more effective, leading to the logarithmic spreading of particles into the empty region. The inset shows the dependence of the displacement of the center of mass on time that has a clear ballistic regime of linear increase with time followed by slow logarithmic growth at later times.
}
\end{figure}

To reproduce correlated hopping in the Hamiltonian~(\ref{Eq:Hr=2}), we introduce two sets of local gates $U_1$ and $U_2$ schematically shown in Fig.~\ref{Fig:Circuit}(a). The first gate, $U_1$, acts on $4$ sites and implements the hopping facilitated by the next nearest neighbor,
\begin{equation}
\label{Eq:U1}
U_1 = \exp\Bigg\{ - \imath \theta\bigg[ \hat{n}_{j}(1-\hat{n}_{j+1})\bigr(c^\dagger_{j+3}c_{j+2} + \text{H.c.}\bigr)\bigg]\Bigg\}.
\end{equation}
The second gate, $U_2$, acts on three sites, and implements the hopping facilitated by the nearest neighbor site:
\begin{equation}
\label{Eq:U2}
U_2 = \exp\Bigg\{ - \imath \theta\bigg[ \hat{n}_{j}\bigr(c^\dagger_{j+2}c_{j+1} + \text{H.c.}\bigr)\bigg]\Bigg\}.
\end{equation}
For a generic choice of the rotation angle $\theta$ these gates cannot be efficiently simulated classically. However, in what follows we fix $\theta$ to the special value, $\theta=\pi/2$, so that gates $U_{1,2}$ map any product state to another product state.
This corresponds to a classical cellular automaton which allows for efficient classical simulation. 
\begin{figure}[h]
\centering
\includegraphics[width=.65\columnwidth]{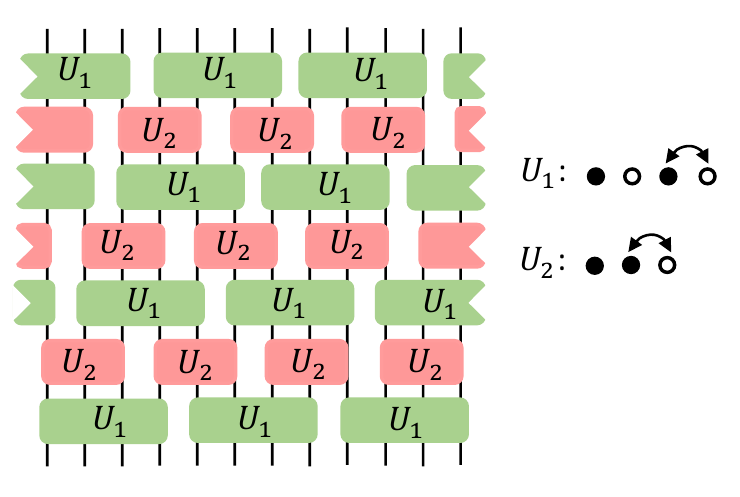}
\caption{\label{Fig:Circuit_sch}
Schematic representation of the circuit used to describe the classical dynamics. The continuous time-evolution $\hat{U}(t)$ is decomposed into a series of $4$-sites gates $U_1$ and of $3$-sites gates $U_2$, whose action is shown on the right part of the Figure.
}
\end{figure}

As each local gate is particle conserving, in order to allow for non-trivial transport, we shift gate position by one site after each layer, as shown in Fig.~\ref{Fig:Circuit}(a). Consequently, the circuit has a $7$-layer unit cell in the time direction. Additionally, the order of gate applications is also important, as the gates $U_{1,2}$ generally do not commute with each other. Alternating the layers of $U_1$ and $U_2$ gates proves to be the best choice, as it implements all allowed particle hopping processes, leading to the circuit shown in Fig.~\ref{Fig:Circuit}(a).

Using this cellular automaton we are able to simulate the time-evolution of very large systems to extremely long times. As the setup implements the same constraint as the Hamiltonian dynamics, we conjecture that it should present similar features. For instance, initializing the system in a dense-empty configuration similar to the $\ket{\text{DW}}$ state, we expect the dense region to spread quickly into the empty one, until eventually it stretches too much and its propagation slows down due to the constraint.

We study the evolution to the domain-wall initial state for a system of $L=298$ sites and $N_p=100$ particles. Since this model is deterministic, the density as a function of circuit depth is a binary function, $n_i(t)\in\{0,1\}$. Figure~\ref{Fig:Circuit}(b) shows the short-time density dynamics ($t<1000$). We observe ballistic particle transport in the dense regime. On the one hand, the position of the rightmost particle moves to the right. On the other hand,  defects (holes) propagate within the dense domain wall state. The simulation reveals notable difference in velocities of holes and spreading of the rightmost particle, that is expected in view of the inversion breaking symmetry within the model. 

The ballistic expansion of the particles is followed by a logarithmic slowdown at later times as shown in Fig.~\ref{Fig:Circuit}(b). Much akin to the Hamiltonian dynamics, this slowdown is due to the lower density reached at later times as the front moves to the right and more particles become temporarily frozen due to the constraint.
To further probe the two distinct behaviors observed in the cellular automaton, in the inset of Fig.~\ref{Fig:Circuit}(b) we show the time-evolution of the displacement of the center of mass $R(t)$ as in Eq.~(\ref{Eq:R tilde}). From the initial linear behavior, $R(t)$ abruptly enters a logarithmic regime as it exceeds the extent of the ballistic region, corresponding to $i\approx180$.

The study of the circuit evolution for the domain-wall initial state then shows the overall similar characteristic inhomogeoneous dynamics as the quantum system. At early times, and close to the initial domain wall $i=N_p$, the transport of particles and holes is ballistic as for $t\leq1$ in the quantum case (see Fig.~\ref{Fig:Rtilde and z}). However, as the density spreads and particle density lowers, ballistic spreading  is replaced by a logarithmic slow dynamics. We notice, however, that the automaton lacks the super-diffusive plateau observed in the Hamiltonian dynamics.

\section{Discussion}\label{Sec:Disc}
In this work, we introduced a family of models characterized by a conserved $U(1)$ charge and strong inversion symmetry breaking. We demonstrate that such models feature recursive quantum Hilbert space fragmentation~\cite{Motrunic2022} that gives rise to weakly entangled eigenstates coexisting with volume-law entangled eigenstates in the spectrum.  In addition, we investigate the dynamics of the system in a quantum quench launched from the domain wall initial state. Although the long time saturation value of particle density is consistent with thermalization, we observe two distinct regimes in particles spreading. An initial superdiffusive particle spreading at high density is dramatically slowed down at lower densities, leading to a logarithmically slow approach of density to its saturation value. We suggest the particular structure of the constrained Hilbert space as a possible explanation of such slow propagation. In addition, we also reproduce the logarithmic dynamics in a classical cellular automaton that features the same symmetries, although at early times the cellular automaton features ballistic dynamics in contrast to slower but still superdiffusive spreading of particles in the Hamiltonian model. 

Our work suggests that the interplay of constraints and broken inversion or other spatial symmetries may lead to new universality classes of weak thermalization breakdown and quantum dynamics. In particular, the quantum Hilbert space fragmentation in the considered model gives rise to a number of weakly entangled eigenstates that can be interpreted as quantum many-body scars~\cite{Serbyn2021,Regnault2022}. The number of these eigenstates scales exponentially with system size. Moreover these eigenstates may be constructed in a recursive fashion, by reusing eigenstates of a smaller number of particles. This is in contrast to the PXP model, where the number of scarred eigenstates is believed to scale polynomially with system size~\cite{Turner2018,Iadecola2019}, though existence of a larger number of special eigenstates was also conjectured~\cite{Ljubotina2022}. 

Although we presented an analytic construction for certain weakly entangled eigenstates and demonstrated their robustness to certain deformations of the Hamiltonian, the complete understanding of quantum recursive Hilbert space fragmentation requires further work. The complete enumeration and understanding of weakly entangled eigenstates may give further insights into their structure and requirements for their existence.  In addition, a systematic study of the emergence of quantum Hilbert space fragmentation in the largest sector of a classically connected Hilbert space in other constrained systems, like the XNOR or the Fredkin models is desirable~\cite{Vasseur2021,Yang2022}. 

From the perspective of particle transport, the numerical data for dynamical exponent that controls particle spreading suggests that constrained models may provide stable generic examples of superdiffusive dynamics~\cite{Ljubotina2017,DeNardis2021,Ilievski2021,Bulchandani2021,Ljubotina2022}. 
This observation differs from results of Ref.~\cite{Vasseur2021} that reported typically slower than diffusive dynamics. This difference may be partially attributed to the fact that Ref.~\cite{Vasseur2021} probed dynamics via the time-evolution of an infinite temperature density matrix that was not projected to the largest connected sector of the Hilbert space. Similarly to quantum Hilbert space fragmentation, our  understanding of transport properties also remains limited. This invites large-scale numerical studies of transport in the particle conserving east models using operator evolution with tensor network methods~\cite{Ljubotina2022}. Such studies would enable accessing much larger system sizes and are likely to provide valuable insights needed for constructing an analytic theory of transport.  In particular, it is interesting to study the dependence of the superdiffusive exponent on the constraint range in the family of models introduced in this work. 

Finally, the models considered in our work may be implemented using quantum simulator platforms. In particular, the Floquet model consists of control-swap gates of various ranges. Thus, an experimental study of such models may reveal novel valuable insights into their physics and the universality of their transport phenomena.

\section*{Acknowledgments}

We would like to thank Raimel A. Medina, Hansveer Singh, and Dmitry Abanin for useful discussions. 
The authors acknowledge support by the European Research Council (ERC) under the European Union's Horizon 2020 research and innovation program (Grant Agreement No.~850899). We acknowledge support by the Erwin Schr\"odinger International Institute for Mathematics and Physics (ESI).

\begin{appendix}

\section{Thermalization within the largest subsector of the Hilbert space}\label{App:therm}

In order to show the ergodic behavior of the eigenstates of the Hamiltonian, we study the distribution $P(s)$ of the energy differences in the sorted eigenspectrum weighted by the mean level spacing $\Delta$, $s_i=(\epsilon_i-\epsilon_{i-1})/\Delta$. It is known that thermal systems which satisfy the eigenstate thermalization hypothesis are characterized by level statistics in agreement with the prediction of the Gaussian orthogonal ensemble (GOE), $P_{\text{GOE}}(s) = \frac{\pi}{2} se^{-\frac{\pi}{4}s^2}.$

\begin{figure}[b!]
\centering
\includegraphics[width=1.01\columnwidth]{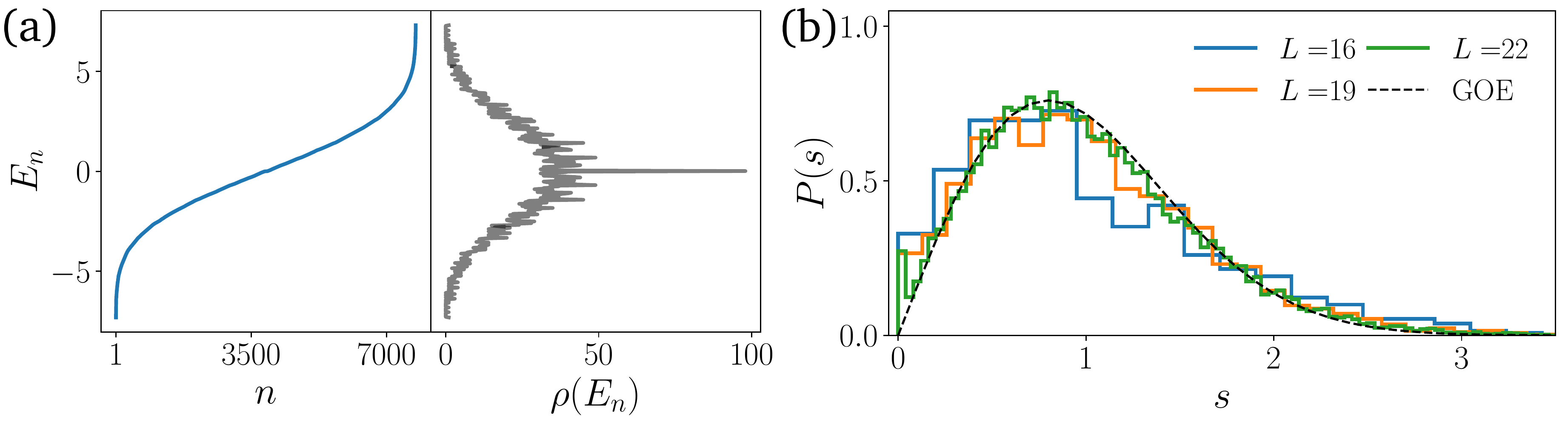}
\caption{\label{Fig:DOS}
(a): As shown in the left sub-panel, the spectrum is symmetric with respect to $E_n=0$, such that for any eigenstate with eigenvalue $E_n$ there is a second state with energy $-E_n$. 
Additionally, the model has a large number of zero energy eigenstates, as highlighted by the peak of the density of states $\rho(E_n)$ in the right sub-panel. We show data for $N_p=7$ and $L=19$. 
(b): The level spacing distribution $P(s)$ shows good agreement with the GOE prediction, shown as a black dashed line, thus confirming the presence of level repulsion within the largest subsector.
}
\end{figure}

However, before discussing the level statistics, the discussion of the density of states is in order. The Hamiltonian $\hat H_2$ has a spectral reflection property with respect to $E=0$ and it presents an exponentially large in system size number of zero modes, as highlighted by the peak in the density of states $\rho(0)$ shown in Figure~\ref{Fig:DOS}(a). The large number of zero energy eigenstates is explained by the bipartite nature of the adjacency graph that describes the Hamiltonian, see Figure~\ref{Fig:G} for an example.
 In a bipartite graph there exist two sets of nodes $\mathcal{P}_{1,2}$ labeled by different product states, such that the action of the Hamiltonian on states belonging to the set $\mathcal{P}_1$ yields a state in the set $\mathcal{P}_2$ and vice versa.
These two partitions are identified by the eigenvalue of the parity operator $\hat{\mathcal{P}} = \prod_j (1-2\hat{n}_j)^j=\prod_j (-\sigma^z_j)^j$, where $\sigma^z_j=2\hat{n}_j-1$ is the corresponding Pauli matrix. It is known that a bipartite graph has a number of zero modes bounded from below by the difference in the size of the two sets $P_{1}$ and $P_2$~\cite{Abrahams1994}. 

In fact, when the two partitions have a different number of states, a non-trivial solution of the Schr\"odinger equation for a zero energy eigenstate can be expressed as a system of $n_1$ linear equations for $n_2$ variables. If $n_2>n_1$, there are at least $n_2-n_1$ linearly independent solutions. In this case, in spite of the bound not being tight, both the number of zero modes and the lower bound from the bipartite structure of the graph describing the Hamiltonian increase exponentially with system size, albeit with different prefactors in the exponent. This suggests that the present understanding of the zero mode subspace is incomplete, inviting further research. In particular, using the disentangling algorithm~\cite{Karle} may give valuable insights.  This may also help to develop a more complete understanding of the recursive Hilbert space fragmentation, since its mechanism relies on the zero energy eigenstates with vanishing particle density on the last sites of the system, see Section~\ref{Sec:Quantum Fragmentation}.

In Figure~\ref{Fig:DOS}(b) we show the level spacing distribution for $L\in[16,22]$ in the interval $[E_\text{GS},-0.1]$, where $E_\text{GS}$ corresponds to the ground state energy. 
Note that due to the spectral reflection property of the Hamiltonian, taking into account only negative energies yields the same results as considering the whole spectrum.
To obtain $P(s)$, we unfold the spectrum in the given interval through polynomial interpolation of the integrated density of states.
The agreement with the GOE prediction suggests that despite the presence of a constraint, the levels develop repulsion within the largest connected sector of the Hilbert space and the model is not integrable. 

\begin{figure}[b!]
\centering
\includegraphics[width=1.01\columnwidth]{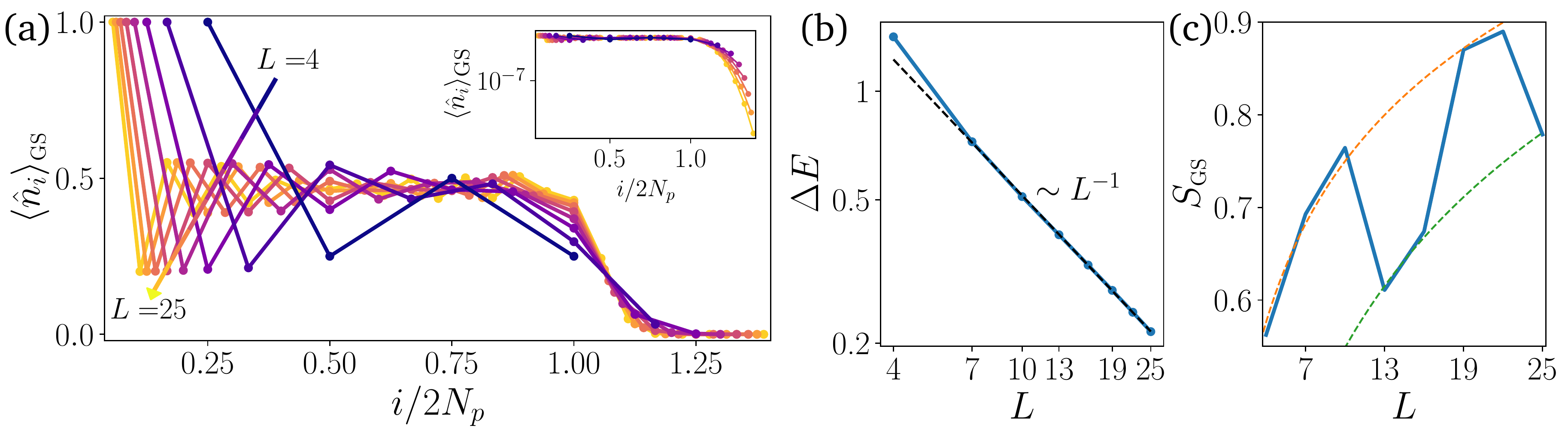}
\caption{\label{Fig:GS}
(a): The density profile of the ground state $\langle \hat{n}_i\rangle_{\text{GS}}$ shows large particle occupation up to $i=2N_p$. Outside this region, the density starts decaying exponentially, as shown in the inset. 
(b):The finite size scaling of the energy gap $\Delta E$ shows that it vanishes as $1/L$, thus indicating that the model is gapless in the thermodynamic limit.
(c):  Entanglement entropy across the central cut grows logarithmically with strong finite size corrections (dashed orange and green lines show logarithmic fits), providing additional evidence that the ground state is critical.
}
\end{figure}

\section{Ground state characterization}\label{App:GS}%

In this Appendix we characterize the ground state, studying the scaling of the energy gap and of the entanglement entropy.
As the Hamiltonian~(\ref{Eq:Hr}) only has hopping terms, the low lying eigenstates need to have a large overlap with product states that maximize a number of configurations to which hopping is allowed to. In graph language, see Figure~\ref{Fig:G} for an example,  these product states correspond to vertices with the largest possible connectivity.
For $r=2$, the state with highest connectivity is $|\underbrace{\bullet\circ\bullet\circ\bullet\dots\bullet\circ}_{2N_p}\underbrace{\circ\circ\circ\dots\circ}_{L-2N_p}\rangle$, with connectivity $2N_p-1$, hence we expect the ground state to have a large weight on the initial $2N_p$ sites. In Figure~\ref{Fig:GS}(a) we plot the density profile of the ground state of the Hamiltonian~(\ref{Eq:Hr=2}) for different system sizes from $L=4$ to $L=25$ against a rescaled $x$-axis $i/2N_p$. The figure confirms the prediction, the ground state is confined within the first $2N_p$ sites, with an exponentially decaying density outside of this region, as shown in the inset. This behavior is different from the one observed in the quantum East model in absence of particle conservation~\cite{Pancotti2020,Marino2022}, where occupation immediately decays exponentially.

We further study the scaling of the energy gap and of the entanglement entropy. As clearly shown in Figure~\ref{Fig:GS}(b), the energy gap $\Delta E$ vanishes as the inverse system size, suggesting that model is in a gapless phase in the thermodynamic limit. Additionally, the entanglement entropy of the ground state across the central cut in the chain presents a slow logarithmic growth. These results suggest that the ground state is critical.

\section{Construction of left parts of separable eigenstates}\label{App:PsiL}

In this section we report the left-restricted eigenvectors entering Eq.~(\ref{Eq:zero S evec}) for all sub-system sizes we were able to investigate numerically for $r=2$. These were used in the main text to correctly count the global number of zero entanglement eigenstates $\mathcal{N}_S$ shown in Figure~\ref{Fig:Special evecs}(b). We remind here that these eigenstates have to fulfill two conditions 
\begin{itemize}
\item[(i)] they have to be an eigenstate on the problem restricted to $m$ particles in $\ell$ sites, with $\ell\leq 3m-2$.
\item[(ii)] They must have zero density on the boundary site $\ell$: $\bra{\psi_m^\ell}\hat{n}_\ell\ket{\psi_m^\ell} = 0$.
\end{itemize}
Additionally we observe that these left-restricted eigenvectors always correspond to zero energy. 

To obtain these states, we take advantage of the large number of zero modes of the Hamiltonian~(\ref{Eq:Hr=2}). Within the degenerate sub-space, one can perform unitary transformations and obtain a new set of zero energy eigenstates where at least one satisfies the condition (ii) above. To find the correct states in an efficient way, we build the matrix $N_{\alpha,\beta} = \bra{E^{m,\ell}_\alpha}\hat{n}_\ell\ket{E^{m,\ell}_\beta}$ of the expectation values  of the density on the last site on eigenstates of the Hamiltonian reduced to $(m,\ell)$. We then diagonalize $N_{\alpha,\beta}$ and check whether it has zero eigenvalues. If so, the corresponding eigenvector is still an eigenstate of the reduced Hamiltonian, and, by construction, it satisfies condition (ii). We notice that this method implements a sufficient condition, which implies that there could be other states that fulfill the same set of restrictions. However, our goal here is merely to provide evidence of existence of these states in several different system sizes.

In the following, we list the states for $m=3,4,5$ and $\ell=6,9,11$ respectively.
\begin{equation}
\begin{split}
\ket{\psi_3^6} &= \frac{1}{\sqrt{2}}\bigr(\ket{\bullet\bullet\circ\circ\bullet\circ}-\ket{\bullet\circ\bullet\bullet\circ\circ}\bigr) \\
\ket{\psi_4^9} &= \frac{1}{2}\bigr(\ket{\bullet\bullet\circ\circ\bullet\circ\circ\bullet\circ} - \ket{\bullet\bullet\bullet\circ\circ\circ\circ\bullet\circ}\bigr) +\frac{1}{4}\bigr(\ket{\bullet\circ\circ\bullet\bullet\bullet\circ\circ\circ}+\ket{\bullet\circ\bullet\bullet\circ\bullet\circ\circ\circ} \\
&+\ket{\bullet\circ\circ\bullet\circ\bullet\bullet\circ\circ} + \ket{\bullet\bullet\bullet\circ\circ\bullet\circ\circ\circ} -\ket{\bullet\bullet\circ\bullet\bullet\circ\circ\circ\circ}-\ket{\bullet\bullet\circ\circ\bullet\bullet\circ\circ\circ} \\
&-\ket{\bullet\circ\circ\bullet\bullet\circ\circ\bullet\circ} -\ket{\bullet\circ\bullet\circ\bullet\circ\bullet\circ\circ} \bigr) \\
\ket{\psi_5^{11}} & = \frac{1}{\sqrt{6}}\bigr(\ket{\bullet\circ\circ\bullet\bullet\bullet\bullet\circ\circ\circ\circ} + \ket{\bullet\circ\bullet\bullet\circ\circ\bullet\bullet\circ\circ\circ} + \ket{\bullet\bullet\circ\circ\bullet\bullet\circ\circ\bullet\circ\circ} + \ket{\bullet\bullet\bullet\circ\circ\circ\bullet\circ\circ\bullet\circ} \\ 
&- \ket{\bullet\circ\bullet\circ\bullet\bullet\circ\bullet\circ\circ\circ} - \ket{\bullet\bullet\circ\bullet\circ\circ\bullet\circ\bullet\circ\circ}\bigr)
\end{split}
\end{equation}

Additional states are present, which we do not write down for the sake of brevity. However, we point out the existence of recursively stacked eigenstates, as mentioned in the main text, and of states where the right part corresponds to a single isolated particle.

\section{Quantum Hilbert space fragmentation for generic Hamiltonian parameters}\label{App:generic r}

Throughout the main text, we often mentioned that the results regarding quantum fragmentation hold irrespective of the range of the constraint $r$ and of the values of the hopping amplitudes $t_\ell$. In the following, we provide evidence in support of the generality of recursive fragmentation.

\begin{figure}[h]
\includegraphics[width=.99\textwidth]{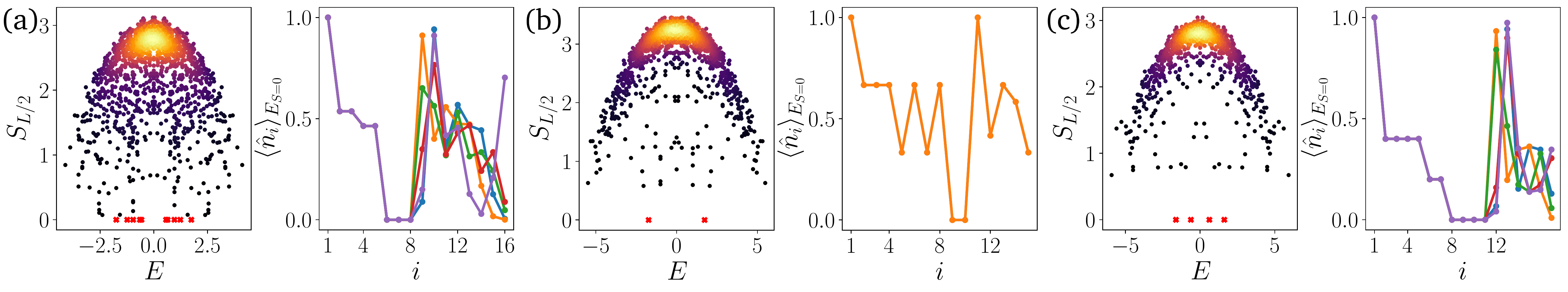}
\caption{\label{Fig:S0 generic}
(a): entanglement entropy of the eigenstates of the Hamiltonian for range $r=2$, random hopping parameters $t_{1}=0.84$, $t_{2}=0.49$ , and system size $L=16$. The presence of zero entanglement eigenstates, highlighted by the red crosses, confirms that quantum fragmentation is insensitive to the value of the hopping amplitudes. (b)-(c): A similar result is obtained for different values of the range $r$. The central panels refer to $r=1$, $N_p=8$ and $L=15$, while the right ones show $r=3$, $N_p=5$ and $L=17$.
}
\end{figure}

In Figure~\ref{Fig:S0 generic}, we first show the entanglement entropy of eigenstates for $r=2$ and Hamiltonian 
\begin{equation}
\label{Eq:Hr=2,t1t2}
\hat{H}  = \sum_{i=2}^{L-1}(t_1\hat{n}_{i-1}+t_2\hat{n}_{i-2} - t_2\hat{n}_{i-1}\hat{n}_{i-2})(\hat{c}^\dagger_{i+1}\hat{c}_i+\text{H.c.}),
\end{equation}
with generic, although homogeneous, hopping amplitudes $t_1,t_2$. In the leftmost panel, we highlight the presence of zero entanglement eigenstates in the half-chain cut for a random choice of the hopping parameters. The density profile of these special eigenstates is similar to the one showed in Figure~\ref{Fig:Special evecs}(a), although the density profile in the left region has more complicated pattern  due to the different values of $t_{1,2}$.

Next, we show the presence of recursive fragmentation in the generic Hamiltonian~(\ref{Eq:Hr}). In the central and right panels of Figure~\ref{Fig:S0 generic} zero entanglement eigenstates (red crosses) appear across the central cut for both $r=1$ and $r=3$. As for the random $t_{1,2}$ case, the structure of these eigenstates is akin to the one obtained in Eq.~(\ref{Eq:zero S evec}), featuring an empty region of $r+1$ sites disconnecting the left region from the right one.
Thus we provide numerical evidence in support of the generic form of the zero entropy eigenstates $\ket{E_{S=0}}$ proposed in the main text. 

\begin{figure}[b!]
\centering
\includegraphics[width=1.01\columnwidth]{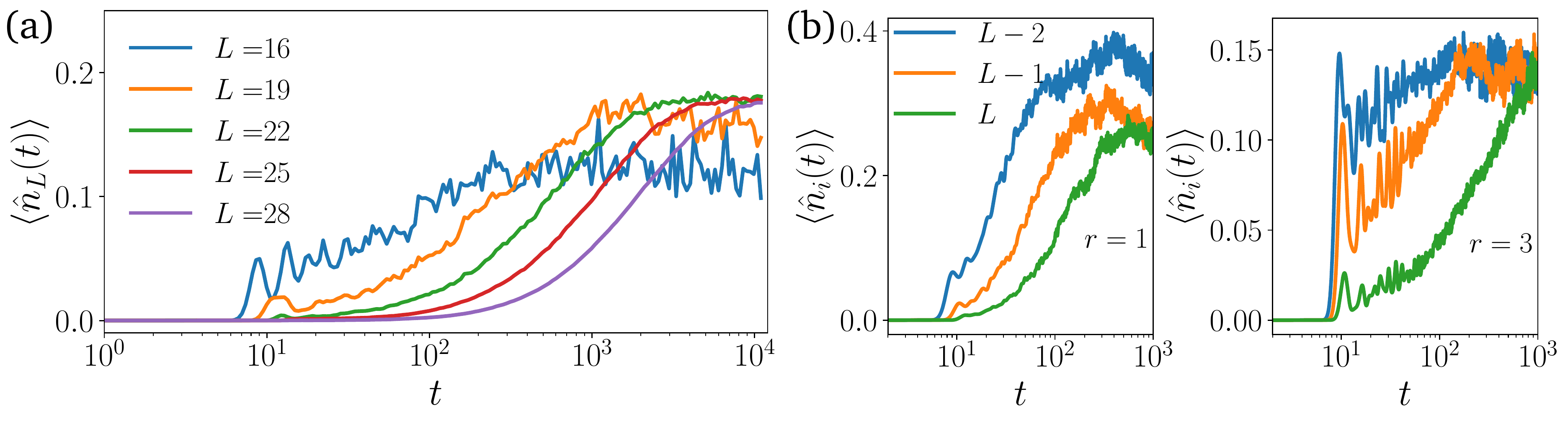}
\caption{\label{Fig:n dynamics scaling-r}
(a): The dynamics of the density on the last site $\langle \hat{n}_L(t)\rangle$ for several different system sizes. The slow logarithmic growth is evident for all $L\geq 16$. At larger system sizes $L\geq19$ the slope becomes independent of system size, as well as the saturation value, thus suggesting a universal behavior. 
(b): The density dynamics for different values of the range $r$ shows always a logarithmic behavior. While the quantitative details change between different values of $r$, the qualitative feature of the logarithmic growth is a constant, thus confirming our claim of generality of the results. The data are obtained on a chain of $N_p=9$ and $L=17$ for $r=1$, and $N_p=6$ and $L=21$ for $r=3$.
}
\end{figure}

\section{Additional evidence of slow dynamics}\label{App:generic dyn}

\begin{figure}[tb]
\centering
\includegraphics[width=1.01\columnwidth]{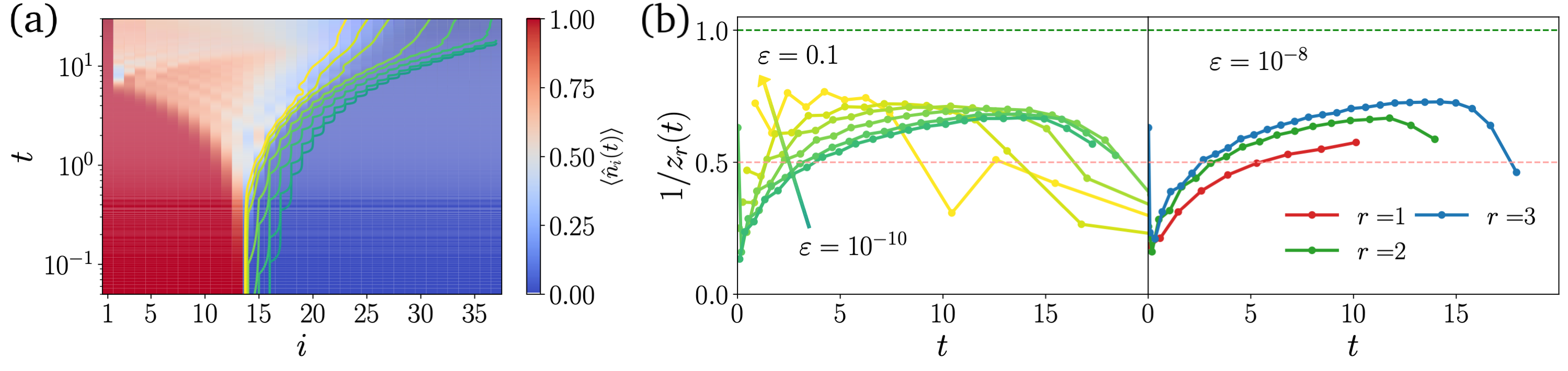}
\caption{\label{Fig:short n dynamics}
(a) Spreading of the density in a system with $L=37$ sites and $N_p=13$ bosons.  Lines of constant value $\varepsilon$ highlight the very different behavior observed in the two regions $i \lessgtr 2N_p$. 
(b) The inverse dynamical exponent $1/z_r(t)$ is always super-diffusive. While for a large threshold it decays to $0$ indicating the onset of logarithmic growth, for small values of $\varepsilon$ the dynamical exponent seems to saturate approaching the asymptotic value (weakly dependent on the threshold value), before the onset of boundary effects. As shown in the right panel, the asymptotic $1/z_r$ is super-diffusive behavior is generic irrespective of the choice of the range of the constraint. The data shown in this panel correspond to $N_p=11$ and $L=21,31,41$ for $r=1,2,3$ respectively.}
\end{figure}

In the main text we provided evidence of slow dynamics from the time-evolution of the density operator in large systems and from the behavior of the root-mean-square displacement. Here, we present some additional data regarding system size scaling of the density dynamics as well as the observation of slow dynamics for generic $r$. Finally, we present an additional measure for the logarithmic behavior of the particles spreading.

In Figure~\ref{Fig:n dynamics scaling-r}(a) we show the system size scaling of the dynamics of the density on the last site of the chain, $\langle \hat{n}_L(t)\rangle$. All the curves present logarithmic growth, and for larger system sizes $L\geq19$ the slope becomes roughly constant. The absence of logarithmic behavior for smaller system sizes $L<16$ is in agreement with the data shown in the main text, where $R(t)$ quickly saturates for $L=13$.

Similar slow dynamics are observed in the time-evolution generated by Hamiltonians with generic constraint range $r$. In Figure~\ref{Fig:n dynamics scaling-r}(b) we present the growth of the density in the last three sites of two chains of length $L=17$ and $L=21$ for $r=1$ and $r=3$ respectively. As the data suggest, the dynamics in the rightmost part of the chain always presents logarithmic behavior, irrespective of the range of the constraint. However, the quantitative details are affected by $r$.

To analyze the spreading of the density, in the main text we presented the behavior of the root-mean-square displacement $R(t)$ together with the respective dynamical exponent $z_R(t)$. Here, we approach the same question using a different measure, namely the time-dependence of the expansion of the density profile. This spreading distance $\delta r$ is defined as the distance from the domain wall boundary, $i=N_p$, at which density becomes larger than a certain threshold $\varepsilon \ll 1$. The spreading distance $\delta r$ is expected to asymptotically behave as a power-law in time, defining a dynamical exponent $z_r$ such that $\delta r \approx t^{1/z_r}$. However, the limited system sizes available to our numerical study do not allow us to reach the asymptotic regime, and we are forced to study the time-dependent analogue $z_r(t)$, obtained through the logarithmic derivative of the spreading distance with respect to time, $(z_r(t))^{-1}=d\ln\delta r /d\ln t$.

In panel~(a) of Figure~\ref{Fig:short n dynamics} we show a heat-map of the density dynamics for $L=37$ sites, superimposed with curves of constant $\langle \hat{n}_i(t)\rangle=\varepsilon$, for values of $\varepsilon \in [0.1,10^{-10}]$, above the accuracy limit $O(10^{-12})$ of the $4$-th order Runge-Kutta algorithm. For each threshold, we show in panel~(b) the time-dependent dynamical exponent. For the largest values of $\varepsilon$ the dynamical exponent has a super-diffusive plateau at $1/z_r(t)\approx 0.7$ before quickly vanishing as expected from the logarithmic dynamics of the density. On the other hand, at smaller thersholds the dynamical exponent seems to saturate to a finite value, before it eventually starts decreasing due to boundary effects.

The saturation value of the time-dependent dynamic exponent for small thresholds has a weak dependence on the value of $\varepsilon$. As $\varepsilon\to0$, $1/z_r$ approaches a $r$-dependent saturation value, monotonically increasing as the range of the constraint becomes larger, as shown in the right panel of Figure~\ref{Fig:short n dynamics}(b). This behavior is in agreement with the expectation that at $r\to \infty$ the system should approach ballistic dynamics.

\end{appendix}

\bibliographystyle{SciPost_bibstyle} 
\bibliography{References.bib}

\nolinenumbers

\end{document}